\documentclass{aastex63}

\shorttitle{GDE measurement by HAWC Data}
\shortauthors{Nayerhoda et al.}

\renewcommand{\baselinestretch}{1.65}
\usepackage{xspace}
\usepackage{float}

\newcommand{\HESS}{H.E.S.S.\xspace}

\newcommand{\degree}{$^\circ$\xspace}

\newcommand{\GP}{Galactic Plane\xspace}
\usepackage{lineno}

\newcommand{\gde}{GDE\xspace}
\usepackage{amsmath}
\usepackage{graphicx}
\usepackage{sidecap}
\usepackage{xcolor}
\newcommand{\amid}{\textcolor{black}}

\usepackage[normalem]{ulem}
\usepackage{soul}
\setstcolor{red}
\begin{document}

\title{Galactic Gamma-Ray Diffuse Emission at TeV energies with HAWC Data}

\collaboration{150}{HAWC Collaboration}


\correspondingauthor{Amid Nayerhoda}
\email{amid.nayerhoda@ifj.edu.pl}
\author{R.~Alfaro}
\affiliation{Instituto de F'{i}sica, Universidad Nacional Autónoma de México, Ciudad de Mexico, Mexico}

\author{C.~Alvarez}
\affiliation{Universidad Autónoma de Chiapas, Tuxtla Gutiérrez, Chiapas, México}

\author{J.C.~Arteaga-Velázquez}
\affiliation{Universidad Michoacana de San Nicolás de Hidalgo, Morelia, Mexico}

\author{K.P.~Arunbabu}
\affiliation{Department of Physics, St. Albert's College (Autonomous), Ernakulam, Cochin, 682018, India}

\author{D.~Avila Rojas}
\affiliation{Instituto de F'{i}sica, Universidad Nacional Autónoma de México, Ciudad de Mexico, Mexico}

\author{R.~Babu}
\affiliation{Department of Physics, Michigan Technological University, Houghton, MI, USA}

\author{V.~Baghmanyan}
\affiliation{Instytut Fizyki Jadrowej im Henryka Niewodniczanskiego Polskiej Akademii Nauk, IFJ-PAN, Krakow, Poland}

\author{E.~Belmont-Moreno}
\affiliation{Instituto de F'{i}sica, Universidad Nacional Autónoma de México, Ciudad de Mexico, Mexico}

\author{C.~Brisbois}
\affiliation{Department of Physics, University of Maryland, College Park, MD, USA}

\author{K.S.~Caballero-Mora}
\affiliation{Universidad Autónoma de Chiapas, Tuxtla Gutiérrez, Chiapas, México}

\author{T.~Capistrán}
\affiliation{Instituto de Astronom'{i}a, Universidad Nacional Autónoma de México, Ciudad de Mexico, Mexico}

\author{A.~Carramiñana}
\affiliation{Instituto Nacional de Astrof'{i}sica, Óptica y Electrónica, Puebla, Mexico}

\author{S.~Casanova}
\affiliation{Instytut Fizyki Jadrowej im Henryka Niewodniczanskiego Polskiej Akademii Nauk, IFJ-PAN, Krakow, Poland}
\affiliation{Max-Planck Institute for Nuclear Physics, Heidelberg, Germany}

\author{O.~Chaparro-Amaro}
\affiliation{Centro de Investigaci'on en Computaci'on, Instituto Polit'ecnico Nacional, M'exico City, M'exico.}

\author{U.~Cotti}
\affiliation{Universidad Michoacana de San Nicolás de Hidalgo, Morelia, Mexico}

\author{J.~Cotzomi}
\affiliation{Facultad de Ciencias F'{i}sico Matemáticas, Benemérita Universidad Autónoma de Puebla, Puebla, Mexico}

\author{S.~Coutiño de León}
\affiliation{Department of Physics, University of Wisconsin-Madison, Madison, WI, USA}

\author{E.~De la Fuente}
\affiliation{Departamento de F'{i}sica, Centro Universitario de Ciencias Exactase Ingenierias, Universidad de Guadalajara, Guadalajara, Mexico}

\author{R.~Diaz Hernandez}
\affiliation{Instituto Nacional de Astrof'{i}sica, Óptica y Electrónica, Puebla, Mexico}

\author{M.A.~DuVernois}
\affiliation{Department of Physics, University of Wisconsin-Madison, Madison, WI, USA}

\author{M.~Durocher}
\affiliation{Physics Division, Los Alamos National Laboratory, Los Alamos, NM, USA}

\author{J.C.~Díaz-Vélez}
\affiliation{Departamento de F'{i}sica, Centro Universitario de Ciencias Exactase Ingenierias, Universidad de Guadalajara, Guadalajara, Mexico}

\author{K.~Engel}
\affiliation{Department of Physics, University of Maryland, College Park, MD, USA}

\author{C.~Espinoza}
\affiliation{Instituto de F'{i}sica, Universidad Nacional Autónoma de México, Ciudad de Mexico, Mexico}

\author{K.L.~Fan}
\affiliation{Department of Physics, University of Maryland, College Park, MD, USA}

\author{N.~Fraija}
\affiliation{Instituto de Astronom'{i}a, Universidad Nacional Autónoma de México, Ciudad de Mexico, Mexico}

\author{A.~Galván-Gámez}
\affiliation{Instituto de Astronom'{i}a, Universidad Nacional Autónoma de México, Ciudad de Mexico, Mexico}

\author{J.A.~García-González}
\affiliation{ITESM}

\author{F.~Garfias}
\affiliation{Instituto de Astronom'{i}a, Universidad Nacional Autónoma de México, Ciudad de Mexico, Mexico}

\author{M.M.~González}
\affiliation{Instituto de Astronom'{i}a, Universidad Nacional Autónoma de México, Ciudad de Mexico, Mexico}

\author{J.A.~Goodman}
\affiliation{Department of Physics, University of Maryland, College Park, MD, USA}

\author{S.~Hernandez}
\affiliation{Instituto de F'{i}sica, Universidad Nacional Autónoma de México, Ciudad de Mexico, Mexico}

\author{B.~Hona}
\affiliation{Department of Physics and Astronomy, University of Utah, Salt Lake City, UT, USA}

\author{D.~Huang}
\affiliation{Department of Physics, Michigan Technological University, Houghton, MI, USA}

\author{F.~Hueyotl-Zahuantitla}
\affiliation{Universidad Autónoma de Chiapas, Tuxtla Gutiérrez, Chiapas, México}

\author{T.B.~Humensky}
\affiliation{Department of Physics, University of Maryland, College Park, MD, USA}

\author{A.~Iriarte}
\affiliation{Instituto de Astronom'{i}a, Universidad Nacional Autónoma de México, Ciudad de Mexico, Mexico}

\author{V.~Joshi}
\affiliation{ECAP}

\author{S.~Kaufmann}
\affiliation{Universidad Politecnica de Pachuca, Pachuca, Hgo, Mexico}

\author{D.~Kieda}
\affiliation{Department of Physics and Astronomy, University of Utah, Salt Lake City, UT, USA}

\author{G.J.~Kunde}
\affiliation{Physics Division, Los Alamos National Laboratory, Los Alamos, NM, USA}

\author{A.~Lara}
\affiliation{Instituto de Geof'{i}sica, Universidad Nacional Autónoma de México, Ciudad de Mexico, Mexico}

\author{H.~León Vargas}
\affiliation{Instituto de F'{i}sica, Universidad Nacional Autónoma de México, Ciudad de Mexico, Mexico}

\author{J.T.~Linnemann}
\affiliation{Department of Physics and Astronomy, Michigan State University, East Lansing, MI, USA}

\author{A.L.~Longinotti}
\affiliation{Instituto de Astronom'{i}a, Universidad Nacional Autónoma de México, Ciudad de Mexico, Mexico}

\author{G.~Luis-Raya}
\affiliation{Universidad Politecnica de Pachuca, Pachuca, Hgo, Mexico}

\author{K. ~Malone}
\affiliation{Space Science and Applications Group, Los Alamos National Laboratory, Los Alamos, NM, USA}

\author{O.~Martinez}
\affiliation{Facultad de Ciencias F'{i}sico Matemáticas, Benemérita Universidad Autónoma de Puebla, Puebla, Mexico}

\author{J.~Martínez-Castro}
\affiliation{Centro de Investigaci'on en Computaci'on, Instituto Polit'ecnico Nacional, M'exico City, M'exico.}

\author{J.A.~Matthews}
\affiliation{Dept of Physics and Astronomy, University of New Mexico, Albuquerque, NM, USA}

\author{P.~Miranda-Romagnoli}
\affiliation{Universidad Autónoma del Estado de Hidalgo, Pachuca, Mexico}

\author{E.~Moreno}
\affiliation{Facultad de Ciencias F'{i}sico Matemáticas, Benemérita Universidad Autónoma de Puebla, Puebla, Mexico}

\author{M.~Mostafá}
\affiliation{Department of Physics, Pennsylvania State University, University Park, PA, USA}

\author[0000-0003-0587-4324]{A.~Nayerhoda}
\affiliation{Instytut Fizyki Jadrowej im Henryka Niewodniczanskiego Polskiej Akademii Nauk, IFJ-PAN, Krakow, Poland}
\affiliation{INFN, Sezione di Bari, Via Amendola 173, Bari, 70126, Italy}

\author{L.~Nellen}
\affiliation{Instituto de Ciencias Nucleares, Universidad Nacional Autónoma de Mexico, Ciudad de Mexico, Mexico}

\author{R.~Noriega-Papaqui}
\affiliation{Universidad Autónoma del Estado de Hidalgo, Pachuca, Mexico}

\author{E.G.~Pérez-Pérez}
\affiliation{Universidad Politecnica de Pachuca, Pachuca, Hgo, Mexico}

\author{D.~Rosa-González}
\affiliation{Instituto Nacional de Astrof'{i}sica, Óptica y Electrónica, Puebla, Mexico}

\author{E.~Ruiz-Velasco}
\affiliation{Max-Planck Institute for Nuclear Physics, 69117 Heidelberg, Germany}

\author{H.~Salazar}
\affiliation{Facultad de Ciencias F'{i}sico Matemáticas, Benemérita Universidad Autónoma de Puebla, Puebla, Mexico}

\author{D.~Salazar-Gallegos}
\affiliation{Department of Physics and Astronomy, Michigan State University, East Lansing, MI, USA}

\author{F.~Salesa Greus}
\affiliation{Instytut Fizyki Jadrowej im Henryka Niewodniczanskiego Polskiej Akademii Nauk, IFJ-PAN, Krakow, Poland}
\affiliation{Instituto de Física Corpuscular, CSIC, Universitat de València, E-46980, Paterna, Valencia, Spain}

\author{A.~Sandoval}
\affiliation{Instituto de F'{i}sica, Universidad Nacional Autónoma de México, Ciudad de Mexico, Mexico}

\author{J.~Serna-Franco}
\affiliation{Instituto de F'{i}sica, Universidad Nacional Autónoma de México, Ciudad de Mexico, Mexico}

\author{A.J.~Smith}
\affiliation{Department of Physics, University of Maryland, College Park, MD, USA}


\author{R.W.~Springer}
\affiliation{Department of Physics and Astronomy, University of Utah, Salt Lake City, UT, USA}

\author{O.~Tibolla}
\affiliation{Universidad Politecnica de Pachuca, Pachuca, Hgo, Mexico}

\author{K.~Tollefson}
\affiliation{Department of Physics and Astronomy, Michigan State University, East Lansing, MI, USA}

\author{I.~Torres}
\affiliation{Instituto Nacional de Astrofísica, Óptica y Electrónica, Puebla, Mexico}

\author{R.~Torres-Escobedo}
\affiliation{Tsung-Dao Lee Institute and School of Physics and Astronomy, Shanghai Jiao Tong University, Shanghai, China}

\author{F.~Ureña-Mena}
\affiliation{Instituto Nacional de Astrof'{i}sica, Óptica y Electrónica, Puebla, Mexico}

\author{L.~Villaseñor}
\affiliation{Facultad de Ciencias F'{i}sico Matemáticas, Benemérita Universidad Autónoma de Puebla, Puebla, Mexico}

\author{E.~Willox}
\affiliation{Department of Physics, University of Maryland, College Park, MD, USA}

\author{H.~Zhou}
\affiliation{SJTU}

\author{C.~de León}
\affiliation{Universidad Michoacana de San Nicolás de Hidalgo, Morelia, Mexico}

\nocollaboration{150}

\author{O. Fornieri}
\affiliation{Gran Sasso Science Institute (GSSI), Viale Francesco Crispi 7, 67100 L’Aquila, Italy}
\affiliation{INFN-Laboratori Nazionali del Gran Sasso (LNGS), Via G.~Acitelli 22, 67100 Assergi (AQ), Italy}

\author{D. Gaggero}
\affiliation{INFN, Sezione di Pisa, Largo B. Pontecorvo 3, 56127 Pisa, Italy}

\author{D. Grasso}
\affiliation{INFN, Sezione di Pisa, Largo B. Pontecorvo 3, Pisa, Italy}

\author{A. Marinelli}
\affiliation{Dipartimento di Fisica ``Ettore Pancini'', Università degli studi di Napoli ``Federico II'', Complesso Univ. Monte S. Angelo, I-80126 Napoli, Italy}
\affiliation{INFN - Sezione di Napoli, Complesso Univ. Monte S. Angelo, I-80126 Napoli, Italy}
\affiliation{INAF - Osservatorio Astronomico di Capodimonte, Napoli, Italy}

\author{S. Ventura}
\affiliation{University of Siena, Siena, Italy}
\affiliation{INFN, Sezione di Pisa, Pisa, Italy}




\begin{abstract}
The Galactic gamma-ray diffuse emission (GDE) is emitted by cosmic rays (CRs), ultra-relativistic protons and electrons, interacting with gas and electromagnetic radiation fields in the interstellar medium. 
Here we present the analysis of TeV diffuse emission from a region of the Galactic Plane over the range in longitude of $l\in[43^\circ,73^\circ]$, using data collected with the High Altitude Water Cherenkov (HAWC) detector. Spectral, longitudinal and latitudinal distributions of the TeV diffuse emission are shown.
The radiation spectrum is compatible with the spectrum of the emission arising from a CR population with an "index" similar to that of the observed CRs.
When comparing with the \texttt{DRAGON} \textit{base model}, the HAWC GDE flux is higher by about a factor of two. Unresolved sources such as pulsar wind nebulae and TeV halos could explain the excess emission.
Finally, deviations of the Galactic CR flux from the locally measured CR flux may additionally explain the difference between the predicted and measured diffuse fluxes.

\end{abstract}



\section{Introduction}   
\label{sec:intro}
Cosmic rays (CRs) are highly energetic hadrons and electrons that fill the Galaxy and carry in the Solar neighbourhood as much energy per unit volume as the average Galactic electromagnetic fields and the thermal and turbulent gas phase, namely about 1 eV/cm$^3$.
CR secondary data suggest that for rigidities around 1 GV cosmic protons and nuclei diffuse in the magnetic fields for a timescale of the order of ${10}^7$ years before escaping the Galaxy \citep{1984acr..book.....B}.
At higher energies, CRs are confined for a shorter time in the Galaxy, as {evidenced} by energy-dependent measurements of the Galactic abundances. 
{The electrically charged nature of cosmic rays does not allow them to travel through space without scattering, absorption, or deflection in the magnetic fields.
Thus these particles redistribute diffusively in the Galaxy contributing to the bulk of Galactic CRs known as background CRs or CR \textit{sea}.}


Direct measurements of the spectra of all CR species have recently reached unprecedented quality \citep{Yoon_2017,Adriani:2011cu,Aguilar2015,Ahn:2010gv,Adriani:2019aft}. Until a few years ago the CR energy spectrum below PeV energies was thought to follow a simple power-law (SPL) with a soft index between -2.7 and -2.8.  A more complex picture, however, has emerged since PAMELA \citep{Adriani:2011cu} and AMS \citep{Aguilar2015} found a hardening \citep{2021PhRvL.126t1102A} in the spectra of CR proton, helium and other primary nuclear species at about 300 ${\rm GeV/n}$ (GeV per nucleon) . 
Above that energy, a combined fit of PAMELA, AMS and CREAM \citep{Ahn:2010gv,Yoon_2017} -- the latter probing energies between 1 {\rm TeV/n} and $10^3$~{\rm TeV/n} (n being the atomic mass number) --  yields a value very close to -2.6 for the proton spectral index \citep{Lipari:2017jou}.  This finding has been confirmed by the CALET experiment which measured the proton spectrum between 50 GeV and 10 TeV \citep{Adriani:2019aft}. 
A softening of the spectrum has been recently revealed by DAMPE above few tens of TeV for both proton and helium spectra \citep{2019SciA....5.3793A,2021PhRvL.126t1102A}.

Since the direct measurements of CR spectra are necessarily restricted to the proximity of the Earth, or for the Voyager measurements \citep{Stone150} to the proximity of the Solar System, the distribution of CRs in other regions of the Galaxy is not known. 
It is strongly debated whether the CR sea is homogeneously distributed 
{ along the Galactic plane} or it shows a gradient towards the inner Galaxy, believed to be produced by a spatial dependence of the  transport properties  \citep{Evoli_2017, Evoli_2008}.
It is also not clear whether the spectral hardening of the primary CR spectra (at 300 GeV/n) measured at Earth is representative of the entire CR Galactic population or it is a local effect, originated by the contribution of one or a few nearby CR sources (see, e.g., \citep{Thoudam:2011aa}). These alternative scenarios can be tested by measuring the diffuse gamma-ray emission from the Galactic Plane from GeV to TeV energies.  

The Galactic diffuse gamma-ray emission originates from the interactions of the background CRs (hadrons and electrons) with the interstellar medium (ISM) gas and interstellar radiation field (ISRF).
CR hadrons interact with matter, producing neutral pions ($\pi^0$), which
in turn decay into gamma rays, while CR electrons produce high-energy gamma-rays via inverse Compton (IC) scattering onto the ISRF photons. At GeV energies, gamma rays are emitted through bremsstrahlung processes when electrons collide with interstellar medium gas. 

So far, the diffuse emission of the Galaxy has been primarily investigated in the tens of MeV to hundreds of GeV energy range with orbital detectors such as COS-B \citep{1988A&A...207....1S},  EGRET\footnote{https://heasarc.gsfc.nasa.gov/docs/cgro/egret/} \citep{1996A&A...308L..21S,1997ApJ...481..205H} and Fermi-Large Area Telescope (LAT)\footnote{https://fermi.gsfc.nasa.gov/}.
The first analysis of diffuse Galactic emission using Fermi-LAT data
generally yielded a good agreement between the local CR spectrum 
and the CR spectrum in the Galaxy \citep{2012ApJ...750....3A}. Recent works  \citep{Acero_2016,Yang2016, Pothast_2018} extract the gamma-ray emissivity in Galactocentric rings and report a hardening and an enhancement of the emissivity in the inner Galactic rings, with a maximum at a distance of $\sim$4 kpc from the Galactic Center.

Measurements of the diffuse emission at TeV energies by HEGRA-IACT have determined an upper limit on the ratio of the diffuse photon flux to the hadronic CR flux \textless  \xspace 2.0$\times10^{-3}$ near the inner Galaxy at 54 TeV \citep{Aharonian:2001ft}. The GDE measured with Milagro at a median energy of 15 TeV for Galactic longitudes between 30\degree and 110\degree and between 136\degree and 216\degree and for Galactic latitudes between -10\degree, and +10\degree is consistent with the predictions of \textit{GALPROP} optimized model everywhere except for the Cygnus region \citep{Abdo_2008}. 
Measurements of the diffuse emission at TeV energies by the ARGO-YBJ detector have yielded a soft gamma-ray spectrum with index -2.9 $\pm$ 0.3 within the region 40\degree \textless  \xspace  $l$ \textless  \xspace 100\degree and $|b|$ \textless  \xspace 5\degree in Galactic longitude and latitude, respectively \citep{Bartoli_2015}. Measurements with the H.E.S.S. instrument favor a significant contribution of $\pi^0$-decay to the total signal of the diffuse emission at TeV energies from the Galactic Plane \citep{Abramowski_2014}. A recent study of the diffuse radiation from the Galaxy at 100 TeV by the Tibet AS-Gamma Collaboration shows that CRs are accelerated beyond PeV energies in our Galaxy and spread over the Galactic disk. The hadronic diffuse component is likely the dominant component of this sub-PeV emission \citep{PhysRevLett.126.141101}. Very recently PeV diffuse radiation has been detected by LHAASO \citep{2023arXiv230505372C}.
 
The High Altitude Water Cherenkov (HAWC) Gamma-Ray Observatory\footnote{https://www.hawc-observatory.org/}, located in Mexico, is well-designed to study CRs and gamma rays at energies between 300 GeV to above 100 TeV. 
The detector observes the Cherenkov lights in the water produced by the charged secondary particles.
Thanks to its large field-of-view, it monitors two-thirds of the sky daily with  $ > 95 \%$ duty cycle \citep{Abeysekara_2017, ABEYSEKARA2023168253}.

{We here present the analysis of the HAWC data above 300 GeV and up to 100 TeV within the region of interest (ROI) spanning $l\in[43^\circ,73^\circ]$ in longitude and $b\in[-5^\circ,5^\circ]$ in latitude (see section \ref{sec:results}).
The spectra and profiles of the \gde are measured and shown within specific portions of the Galactic Plane included in ROI (namely $b\in[-2^\circ,2^\circ]$, $b\in[-4^\circ,4^\circ]$, and $b\in[-5^\circ,5^\circ]$).} Future analysis will extend the range in longitude to cover the entire part of the sky visible to HAWC. 
Finally, we compare the measured gamma-ray diffuse emission spectra as well as its latitudinal and longitudinal profiles with those obtained with a reference CR transport model implemented with the \texttt{DRAGON} code \citep{Evoli_2008, Evoli_2017}.

The paper is organized as follows. The analysis of the diffuse emission from the Galactic Plane is described in details in section \ref{sec:method}. Section \ref{sec:results} presents, discusses, and compares the HAWC results with other measurements. Section \ref{sec:conclusions} contains the conclusions.
{The differential flux profiles are presented in Appendix \ref{sec:DFP}, and the comparative analysis between ARGO-YBJ measurements and \texttt{DRAGON} prediction is delineated in Appendix \ref{ARGO-DRAGON}.}
The \texttt{DRAGON} code is also introduced in Appendix \ref{sec:dragon}.

\section{Analysis}  \label{sec:method}
 
\subsection{Data Set}
The analysis presented in this study is based on a data set accumulated over 1,347 days by HAWC from 2015 to 2019. 
Standard reconstructed data generated by the HAWC production of Pass 4 \citep{2017ApJ...843...40A} is used. The data are categorized into 9 energy bins corresponding to the fraction $f_{hit}$ (fractional hit) of PMTs triggered based on the sum of the active PMTs \citep{Abeysekara_2017}.
In Table \ref{tab:bins}, the $\tilde{E}_\gamma^\mathrm{MC}$ column represents the median energy of the simulated gamma-ray photons in the different analysis bins for a Crab-Nebula-like source (dec$=20^\circ$ and for an energy spectrum $E^{-2.63}$). 
By choosing bins $1 - 9$  in the analysis, an energy range between 300 GeV to more than 100 TeV, \citep{2017ApJ...843...40A} is considered.
   
\begin{table}
\small
\centering  
\begin{tabular}{p{2cm}  p{3cm}  p{3cm}  p{1cm} }
 \hline
 \multicolumn{4}{c}{Properties of the analysis bins} \\
 \hline
$\mathcal{B}$ &$f_\mathrm{hit}$ &$\psi_{68}$ & $\tilde{E}_\gamma^\mathrm{MC}$ \\
&(\%) &(\degree) & (TeV) \\
 \hline
      1 &  6.7 --  10.5 & 1.03 &  0.7 \\
      2 & 10.5 --  16.2 & 0.69 &  1.1 \\
      3 & 16.2 --  24.7 & 0.50 &  1.8 \\
      4 & 24.7 --  35.6 & 0.39 &  3.5 \\
      5 & 35.6 --  48.5 & 0.30 &  5.6 \\
      6 & 48.5 --  61.8 & 0.28 &   12 \\
      7 & 61.8 --  74.0 & 0.22 &   15 \\
      8 & 74.0 --  84.0 & 0.20 &   21 \\
      9 & 84.0 -- 100.0 & 0.17 &   51 \\
\hline
  \end{tabular}
\caption{Properties of the 9 analysis bins; bin number: $\mathcal{B}$, event size: $f_\mathrm{hit}$, 68\% PSF containment: $\psi_{68}$, and median energy for a reference  source of spectral index: $-2.63$ at a declination of 20\degree $\tilde{E}_\gamma^\mathrm{MC}$ \citep{2017ApJ...843...40A}.}
    \label{tab:bins}
\end{table}
  
\subsection{Analysis Regions} \label{regions}
The analysis of the diffuse emission reported here is focused on a region of interest (ROI) restricted to a portion of the \GP within the intervals $l\in[43^\circ,73^\circ]$ in longitude and $b\in[-5^\circ,5^\circ]$ in latitude, as defined in the Galactic coordinate system. { { The analysis is carried out by dividing the longitudinal range of the ROI into three sub-regions, namely $l\in[43^\circ,56^\circ]$, $l\in[56^\circ,64^\circ]$ and
$l\in[64^\circ,73^\circ]$, each region being defined in such a way that}}
\amid{no significance excess larger than five ($\sigma$ \textgreater 5)}
is located on the borders of the regions. Hence, none of the regions shares a source. 

\subsection{Event and Background Maps}

In HAWC the data analysis is based on the production of maps of the events and background. The background is estimated with a method known as ``direct integration'', which is used to fit the isotropic distribution of events that pass the gamma-ray event selection \citep{Atkins_2003}. The event maps are simple histograms of the reconstructed events $\lambda_{ij} = b_{ij} + \sum_{k}\gamma_{ijk}$, \amid{that pass gamma/hadron cuts \citep{2017ApJ...843...40A}}, where $\lambda_{ij}$ is the event count in the $j_{th}$ pixel of the $i_{th}$ $f_{hit}$ bin, $b_{ij}$ is the background events in the $j_{th}$ pixel of the $i_{th}$ \textit{$f_{hit}$} bin, and $\gamma_{ijk}$ is the expected number of gamma rays from the $k_{th}$ source in the $j_{th}$ pixel of the $i_{th}$ \textit{$f_{hit}$} bin  (Data = background + signal).

For the present analysis, the maps are produced using HEALPix pixelization  
\citep{2001A&A...376..359H}. The map pixelization is performed with Nside = 1024 for a mean spacing between pixel centers of about 0.05\degree, which is small compared to the typical point spread function (PSF) of the reconstructed events as shown in Table \ref{tab:bins} (the $\psi_{68}$ column represents the 68\% containment angle of the PSF, for a source similar to the Crab Nebula).

\begin{figure}[htbp]
\begin{center}
\includegraphics[width=18cm]{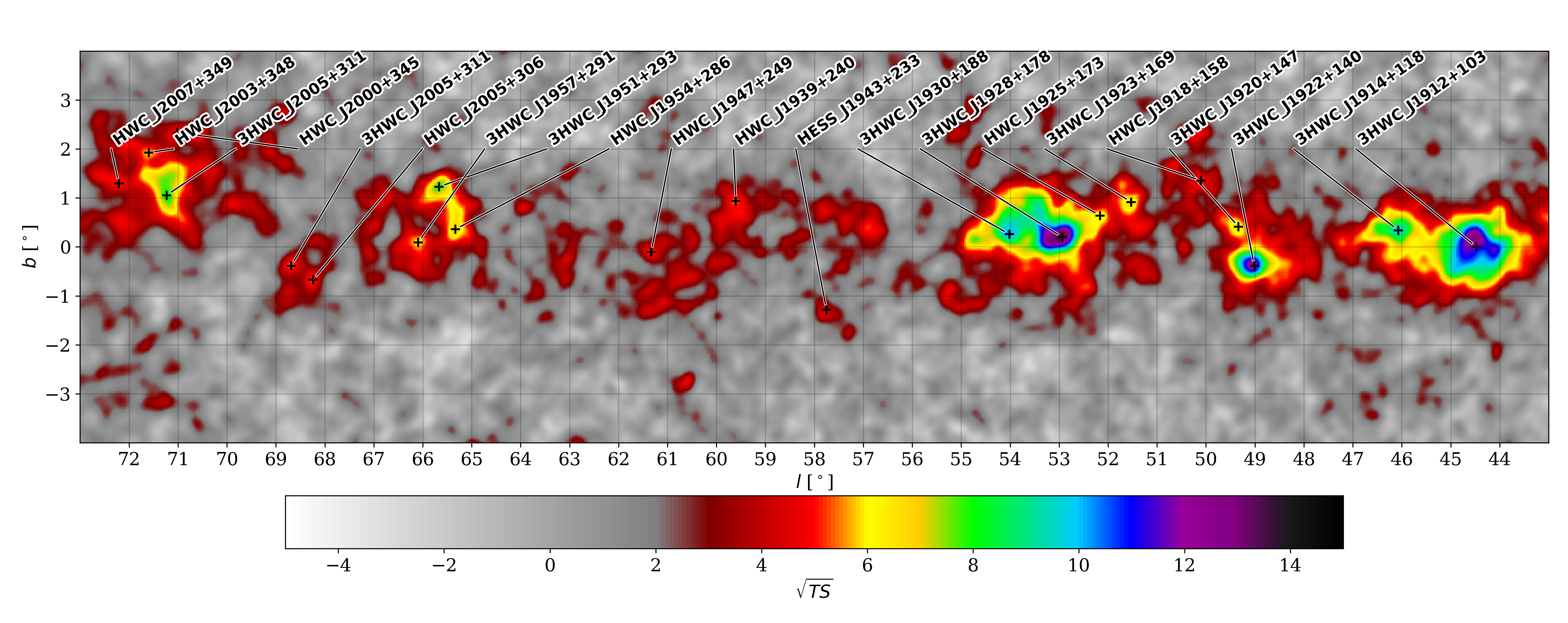}
\end{center}
\caption{Significance map of the total emission measured by HAWC (``original map'') shown over a range in latitude between {$b\in[-4^\circ,4^\circ]$. A significance map is a 2-D visualization of the significance value per each pixel, where significance = $\sqrt{TS}$ \citep{wilks1938large}, and TS is the Test Statistic, as defined using the likelihood
ratio \citep{2017ApJ...843...40A}.}}
   \label{sig-org}
\end{figure}

\begin{figure}
\centering
\includegraphics[width=12cm]{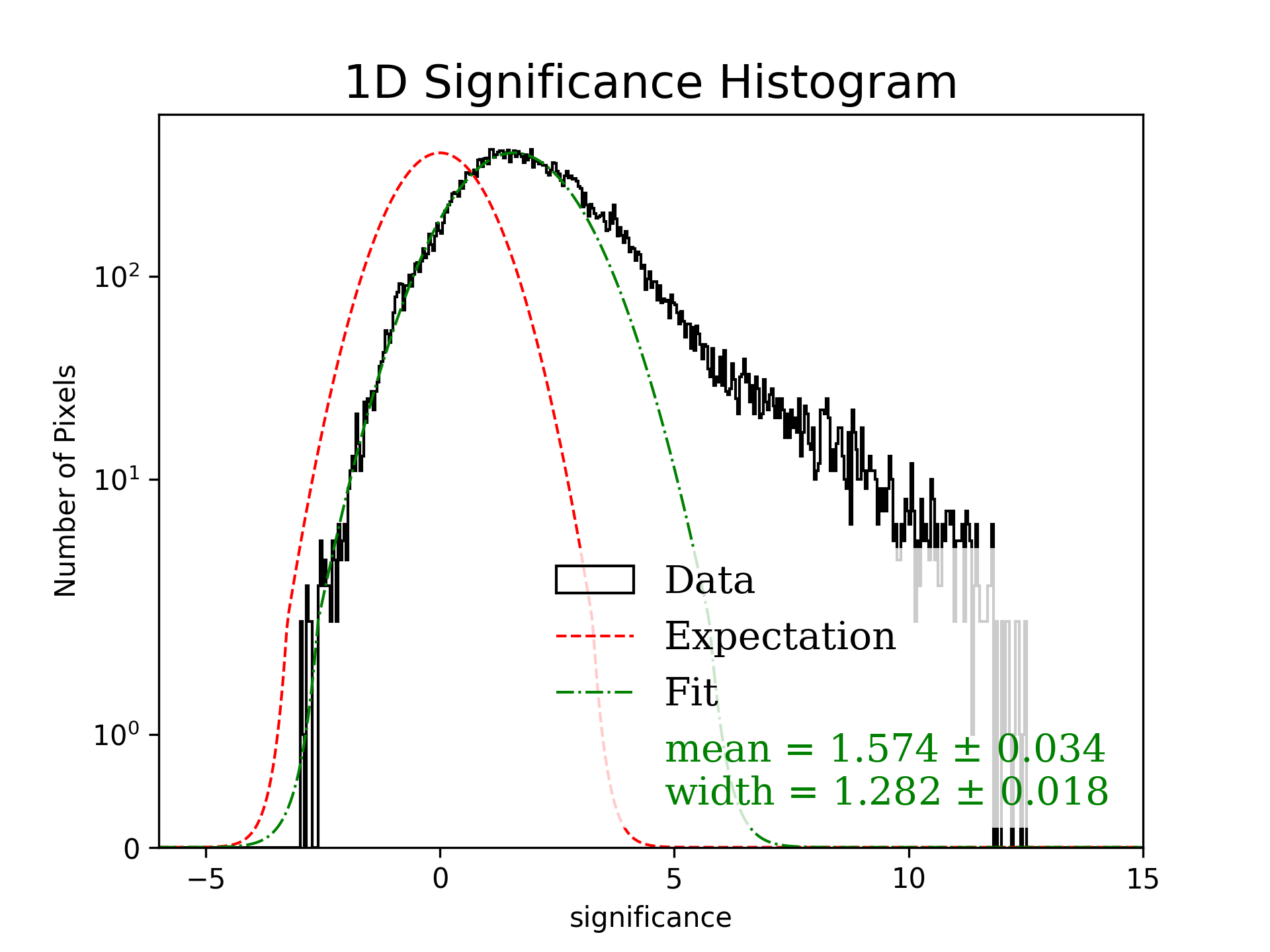}
\includegraphics[width=12cm]{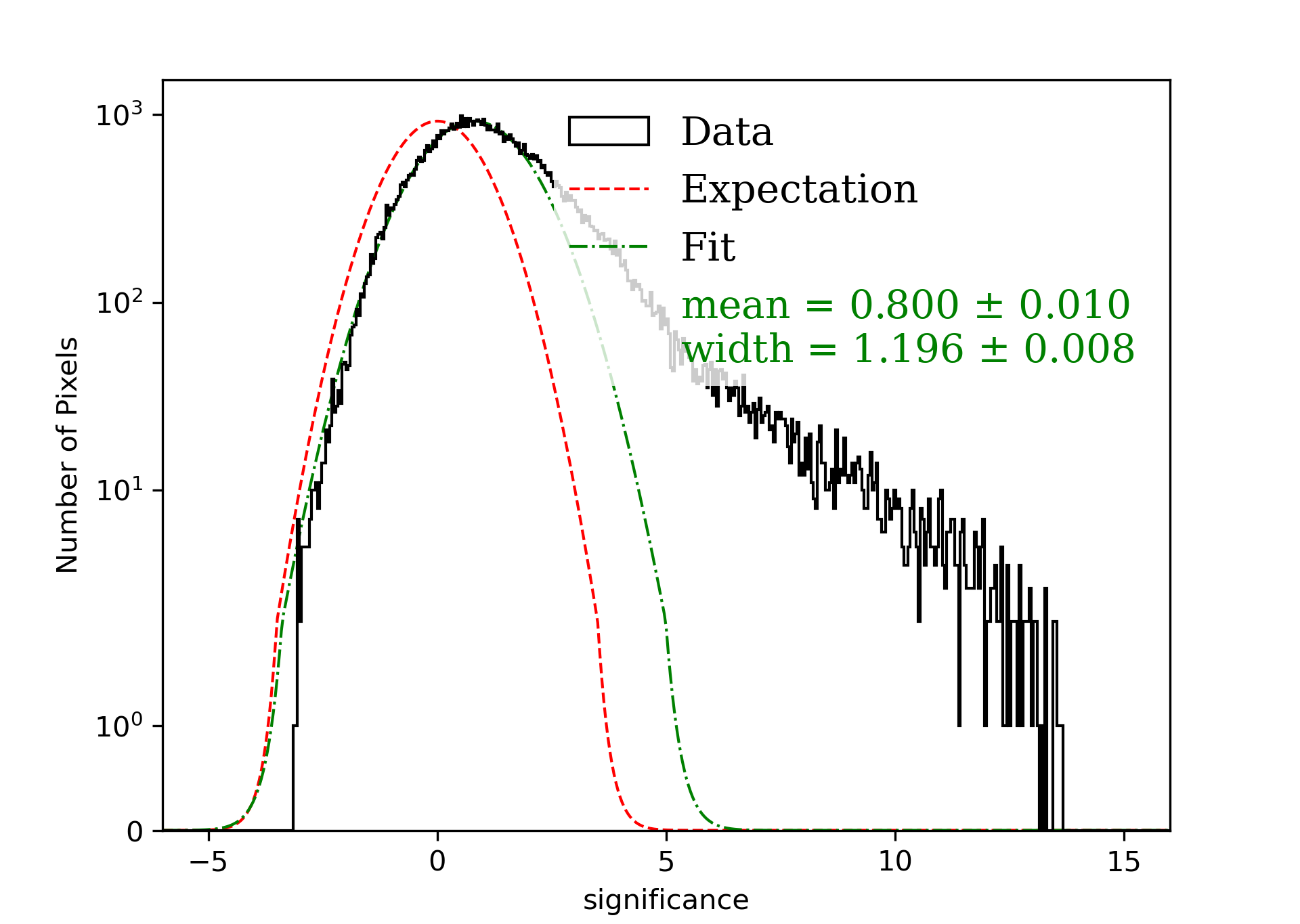}
\caption{Significance histograms of HAWC data in the region restricted in $l\in[43^\circ,73^\circ]$ and $b\in[-2^\circ,2^\circ]$ (top) {and in $l\in[43^\circ,73^\circ]$ and $b\in[-4^\circ,4^\circ]$} (bottom) . The red dashed line represents a standard normal distribution and the green line is a Gaussian fit of the significance of the given data. 
There is an excess with significance {greater} than $\sim$5 in the region considered.}
\label{org_hist}
\end{figure}

In Figure \ref{sig-org}, the significance map\footnote{
The maps are computed using a SPL model with a spectral index of -2.7 for a point source search map.
} of the data measured by HAWC from the Galactic Plane (hereafter called ``original map'') is shown, and the corresponding 1D significance histogram is presented in Figure \ref{org_hist}. 
{We do not show the maps  beyond {$b\in[-4^\circ,4^\circ]$ as the radiation is not significant at higher latitudes.}}

\subsection{Analysis of the GDE} \label{process}
The radiation measured by HAWC from the Galactic Plane, shown in Figure \ref{sig-org} (``original map''), is the sum of different contributions: GDE, produced by background CR protons and electrons, and emission from point-like and extended gamma-ray sources. The total flux measured by HAWC from the Galactic Plane, $F_{tot}$, can thus be written as:
\begin{equation}
F_{\text{tot}} 
= {F_{\text{sources}}} + F_{\text{{GDE}}} 
\label{spectrum}
\end{equation}
\noindent
where ${F_{\text{sources}}}$ is the flux produced by all sources (${\Sigma}_{i}$ ${F_{{source}_i}}$), and $F_{GDE}$ is the flux corresponding to the GDE. 
The ``original map'' is the starting point of the analysis of the GDE, which will be carried out essentially in two steps. In the first step, a map of the source-emission will be obtained and then subtracted from the measured HAWC map (``original map'', Figure \ref{sig-org}). In this way, a ``source-subtracted'' map, which in principle should contain only GDE radiation, will be generated. In the second step, the analysis of the source-subtracted map will yield the spectral, longitudinal, and latitudinal features of the GDE emission.

\begin{figure}[htbp]
\centering
\includegraphics[width=18cm]{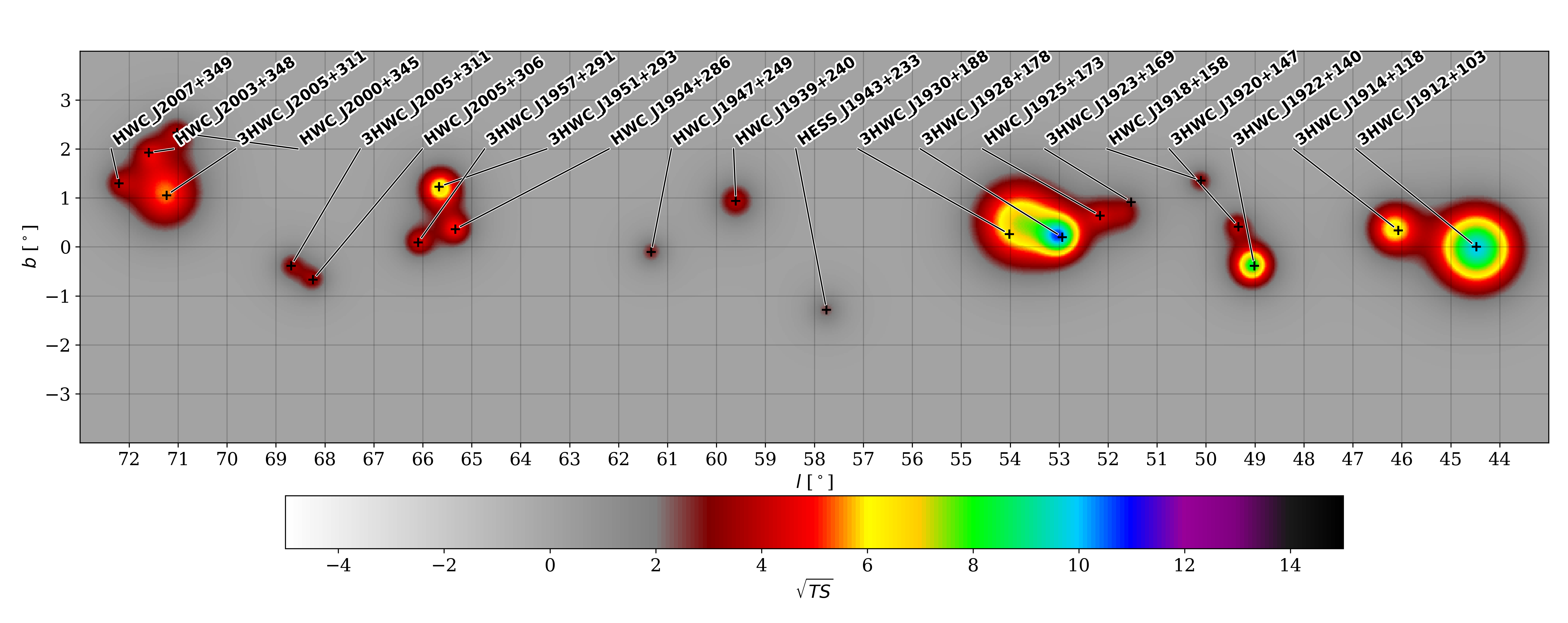}
\includegraphics[width=18cm]{ 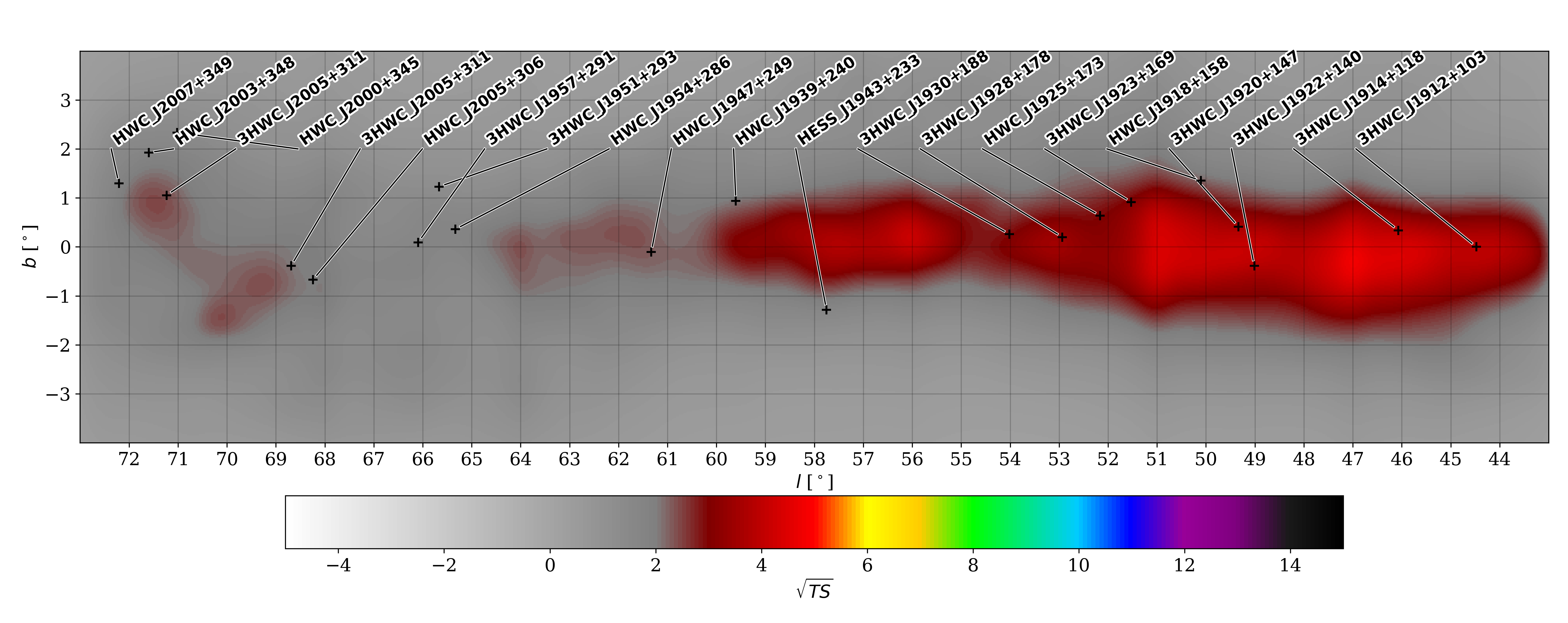}
\includegraphics[width=18cm]{ 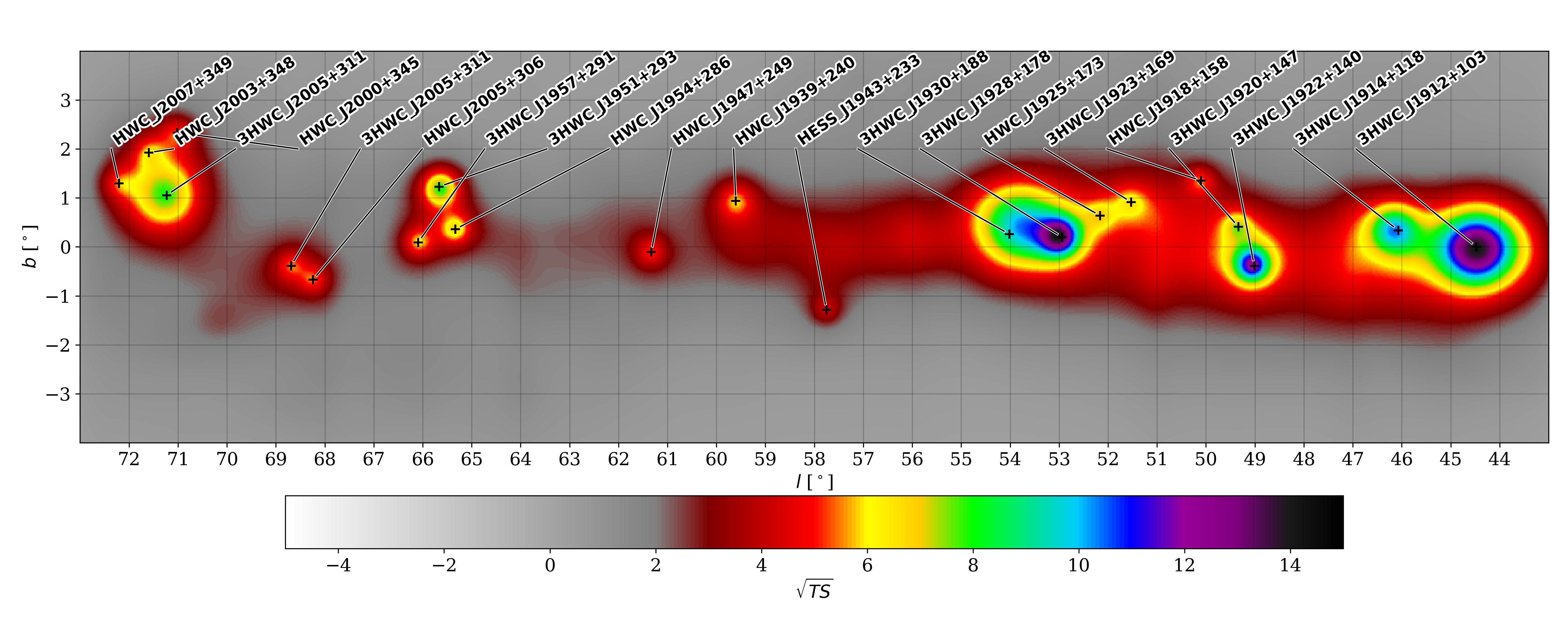}
\caption{
Significance maps of the reference models. 
Top: the fitted model for point-like and extended sources used to obtain the source-subtracted map.
Middle: Reference model of \gde obtained from \texttt{DRAGON}.
Bottom: Sum of the fitted model for the sources and \gde.}
\label{models}
\end{figure}

The contribution of source emission to the total emission is obtained by a multiple-source fitting process, in which a model for the total radiation, including point-like or extended gamma-ray sources resolved by HAWC and the GDE (treated as an extended source), is fitted to the ``original map''.\\
The best fit for the hotspots --- characterized as significance excess larger than three ($\sigma$\textgreater 3) --- in the ``original map'', results in 21 (either point-like or extended) sources, as labeled in Figure \ref{models} \footnote{Spectral parameters and extension (in the case of extended sources) of sources labeled in significance maps differ from those in the catalogs (\HESS \citep{hess-2018}, HAWC \citep{albert20203hwc}), since they are based on a model fitting including a GDE model in the analysis.}

\begin{figure}[htbp]
\begin{center}
   \includegraphics[width=18cm]{ 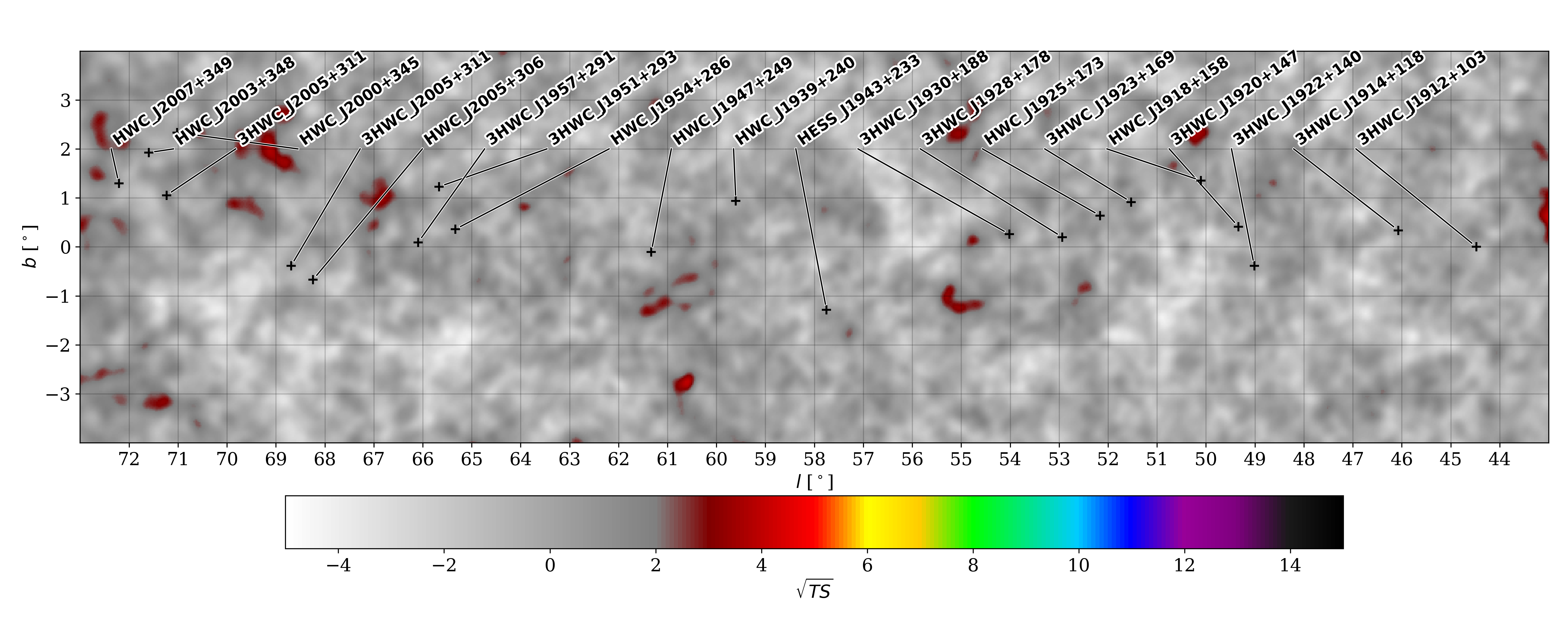}
\end{center}
\caption{Significance of the residual map, created by subtracting the model of the source and \gde (Figure \ref{models}, bottom), obtained in the multi-source fitting procedure, from the ``original map'' (Figure \ref{sig-org}).}
\label{res_GS}
\end{figure}

{ The present study proposes a model in which the spectral distributions of the 21 sources are assumed to follow power-law spectra with exponential cutoffs, while the spectral distribution for the diffuse emission is represented by a simple power-law (SPL). Furthermore, the morphological shape of extended sources in the model is approximated by a Gaussian distribution.
To incorporate the morphological features of the GDE, a 2-dimensional morphological template is included in the model, which is obtained by summing the contributions of $\pi^0$-decay and Inverse Compton (IC) from the DRAGON code (see Appendix \ref{sec:dragon}).

The spectral parameters of both the GDE and the sources, as well as the size of extended sources, are considered as free parameters allowed to vary and be fitted. The fitting procedure is performed using a likelihood method based on the Multi-Mission Maximum Likelihood (3ML) framework \citep{vianello2015multimission}. The objective is to identify the best-fitting model that characterizes the spectral and morphological properties of all 21 sources in the Galactic Plane.
To assess the optimal model, a residual map and significance histogram are obtained by subtracting the model for sources and GDE (Figure \ref{models}, bottom) from the ``original map''. This procedure is repeated many times until the optimal model, for which the residual map shows no significant additional hotspots is obtained. The residual or background map is shown in Figure \ref{res_GS}.} 

The multi-source fitting yields thus the best possible fit for the spectrum and morphology of all sources and therefore, the total flux due to all sources (${F_{\text{sources}}}$). 
Figure \ref{models} (top) shows the map of the fitted source flux. The GDE reference model used in the fitting procedure is plotted in the middle panel, while the model of GDE + sources is shown in the bottom panel.

\begin{figure}
\centering 
\includegraphics[width=12 cm]{ 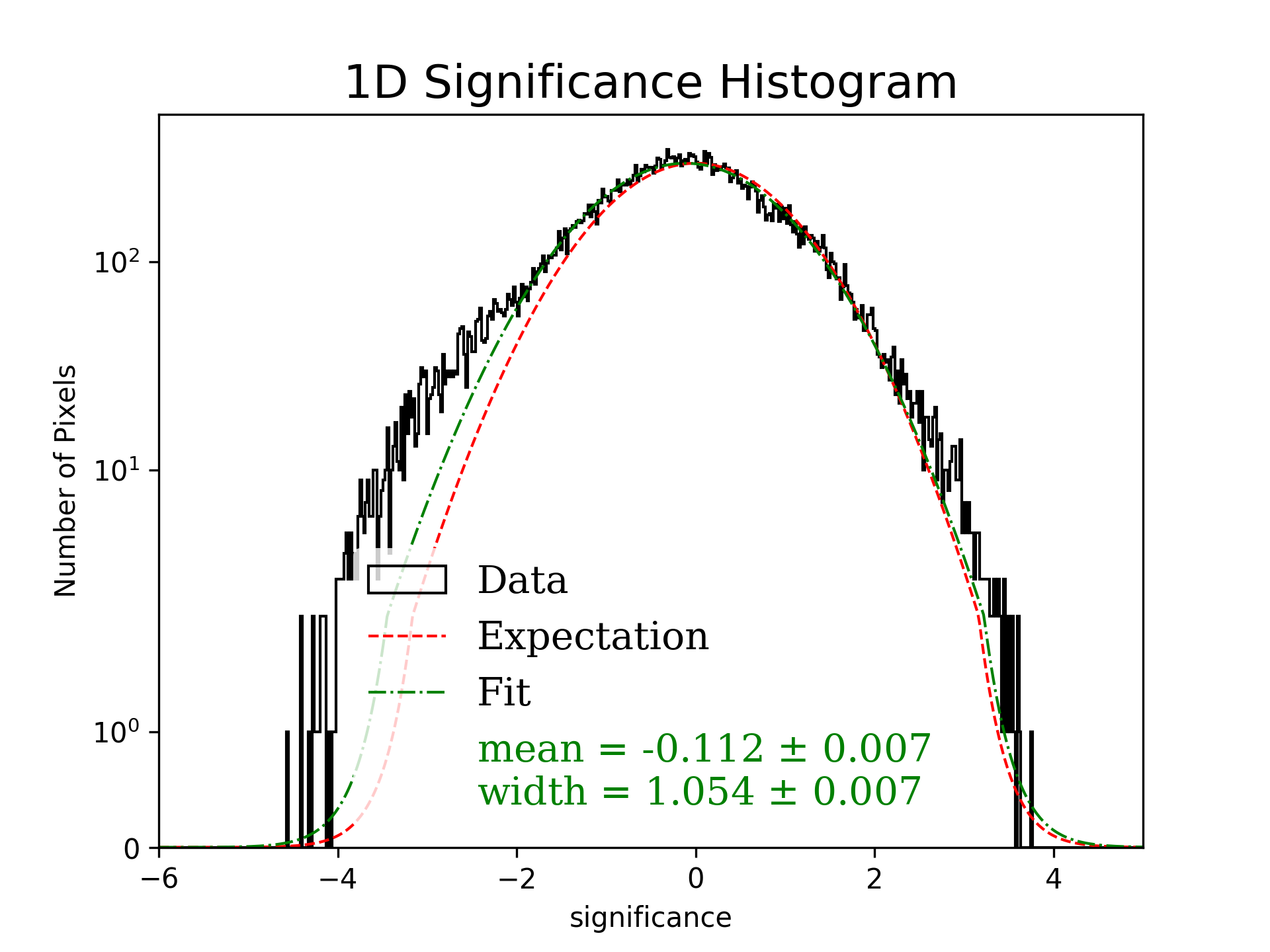}
\includegraphics[width=12 cm]{ 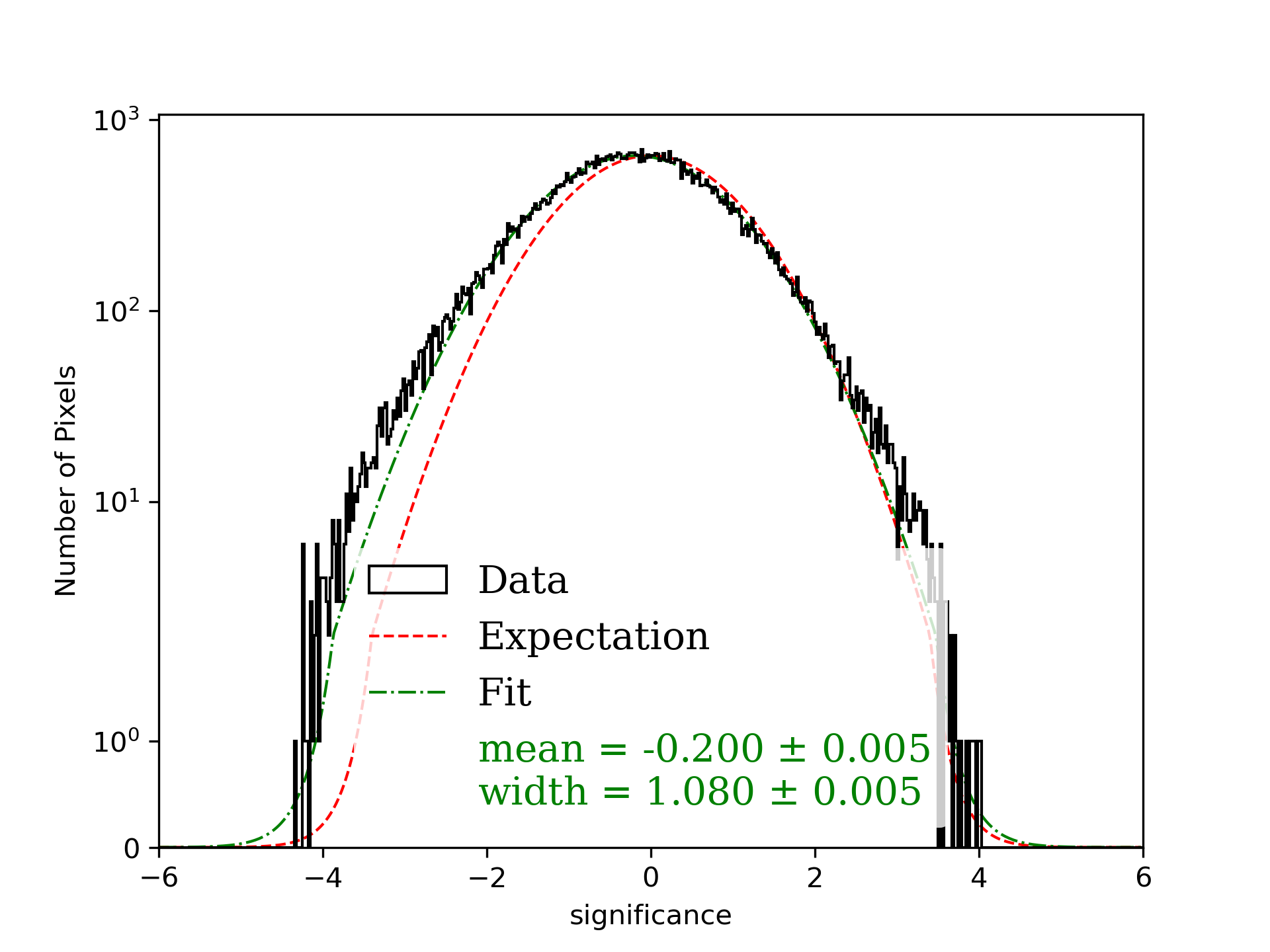}

\caption{Significance histograms of residual map in the region restricted in $l\in[43^\circ,73^\circ]$ and $b\in[-2^\circ,2^\circ]$ (top) {and in $l\in[43^\circ,73^\circ]$ and $b\in[-4^\circ,4^\circ]$}(bottom). The red dashed line represents a standard normal distribution and the green line is a fit of the significance of the given data.}

   \label{res_hist}
\end{figure}

In Figure \ref{res_hist} the significance histogram of the residual map in the region restricted to $b\in[-2^\circ,2^\circ]$ {and to $b\in[-4^\circ,4^\circ]$} is presented. 
The residual map, which shows no 
excess above 5 sigmas, helps test the goodness of the models assumed for sources and GDE.

\begin{figure}[htbp]
  \centering
\includegraphics[width=18cm]{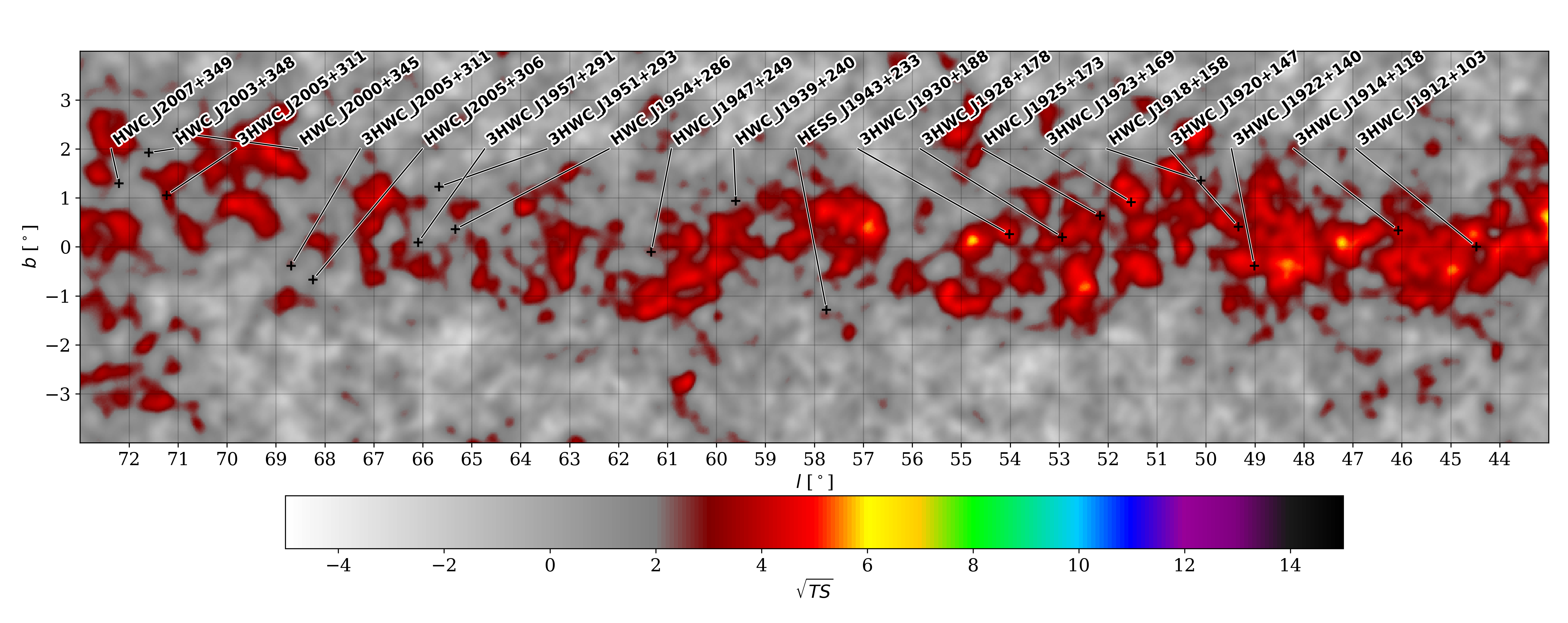}
\caption{Significance map of the source-subtracted map, generated by subtracting the model of the source (Figure \ref{models}, top), obtained in the multi-source fitting procedure, from the ``original map'' (Figure \ref{sig-org}).}
\label{res_G}
\end{figure}

We subtract the model of the sources in Figure \ref{models} (top) from the ``original map'' (Figure \ref{sig-org}) to obtain a ``source-subtracted map'' (GDE map) shown in Figure \ref{res_G}, ($F_{\text{{GDE}}} = F_{\text{{tot}}} - F_{\text{{sources}}}$), which is used to determine the GDE spectral
properties in the final step of the analysis.

\section{Results} \label{sec:results}

\subsection{Galactic Diffuse Emission Spectrum}
 The \gde distribution is reported in Table \ref{tab:GDE} for the regions of the \GP over latitude ranges of $|b|$ \textless  \xspace 2\degree and $|b|$ \textless \xspace 4\degree.
The contribution of the GDE flux with respect to the total flux measured with HAWC is computed for two energy ranges and reported as f$_{10}$ and f$_{100}$\footnote{ $ f_x =\frac{ \int_{0.3(TeV)}^x   F_{GDE}}{    \int_{0.3 (TeV)}^x  F_{tot}} $,  where $F_{\text{{GDE}}}$ and $F_{\text{{tot}}}$ are GDE flux and total HAWC flux respectively.}. f$_{10}$ is the fractional contribution of the GDE flux to the total flux for energies between 300 GeV and 10 TeV, while f$_{100}$ is the fractional contribution of the GDE flux to the total flux up to 100 TeV.

As shown in Table \ref{tab:GDE} the contribution of the \gde flux (both f$_{10}$ and f$_{100}$) varies between 67.2\% and 88.1\%, which implies that the \gde contributes significantly to the total emission of TeV gamma-rays from the Galactic Plane. 
The average GDE flux {(per sr)} over latitudes $|b|$ \textless  \xspace 2\degree is greater than the average GDE flux  in the broader regions with $|b|$ \textless  \xspace 4\degree. In the regions closer to the center of the Galactic disk (b = 0), the \gde is in fact expected to be brighter (see Figures \ref{lat-profile-flux}). 
{{ On the other hand both fractional contributions, f$_{10}$ and f$_{100}$, show that the GDE is less dominant for $|b|$ \textless  \xspace 2\degree (see Table \ref{tab:GDE}). In the outer Galaxy the emission from identified sources contributes less, as expected.}}

\begin{table} 
\small 
\centering
\begin{tabular}{p{1cm} p{1cm} p{1cm}| p{4cm}  p{4cm} ||p{1.5cm} p{1.cm}}
 \hline
 $l_{min}$&$l_{max}$ &  $|b|$  \textless  \xspace & $F_7$ $\times {10^{-12}}$ & Index &f$_{10}$&f$_{100}$ \\
($^\circ$) &($^\circ$) &($^\circ$) &(TeV{$^{-1}$} s$^{-1}$cm$^{-2}$sr$^{-1}$)  & &\%&\% \\
\hline 
\hline 
    43 &  73&2& 8.89 $\pm$ 0.37$^{-0.70}_{+0.48}$& -2.61 $\pm$ 0.03$^{-0.04}_{+0.02}$ &72.7 &71.8\\
    43& 73 & 4& 5.45 $\pm$  0.25$^{-0.44}_{+0.38}$&-2.60 $\pm$ 0.03$^{-0.04}_{+0.01}$ &76.1&75.3 \\
\hline
\hline 
    43 & 56 &2& 9.9 $\pm$  0.6&-2.70 $\pm$ 0.04 &68.8&67.4\\
    43 & 56 &4& 5.8 $\pm$  0.4&-2.69 $\pm$ 0.05  &73.1&71.7  \\
\hline
    56 & 64 &2& 8.9 $\pm$  0.7&-2.58 $\pm$ 0.06&86.4&86.6  \\
    56 & 64& 4&5.2 $\pm$  0.5&-2.60 $\pm$ 0.07 &87.9&88.1  \\
\hline
    64 & 73 &2& 7.8 $\pm$  0.7&-2.48 $\pm$ 0.07&67.2&67.2 \\
    64 & 73 &4& 5.5 $\pm$  0.45&-2.51 $\pm$ 0.06&73.7&73.4 \\
\hline
 \end{tabular}
\caption{Spectrum of the \gde in various sub-regions of the ROI. The first error represents the statistical; the second shows the systematic uncertainty.
For each region, the \gde parameters are reported for $|b|$ \textless  \xspace  2\degree and  $|b|$ \textless  \xspace 4\degree. f$_{10}$  and f$_{100}$  are the fraction of the \gde flux with respect to the total flux, up to 10 TeV and 100 TeV respectively. The flux $F_7$: differential flux at 7 TeV.
The same sources of systematic uncertainty considered in HAWC’s performance papers \citep{Abeysekara_2017c},\citep{Abeysekara_2019} are considered here.}
\label{tab:GDE}
\end{table}

The average spectral index over the region  43\degree  \textless  \xspace $l$  \textless  \xspace 73\degree varies from $-2.61 \pm 0.03$ for latitudes $|b|$ \textless  \xspace2\degree to $-2.60\pm 0.03$ $|b|$ \textless  \xspace4\degree. 
Assuming that the GDE is mostly contributed by hadronic interactions, from the spectral feature of the GDE, we can deduce the spectral feature of the CR population producing this emission. The spectral index of the Galactic background CR population over an energy range between 1-$10^3$~{\rm TeV}, which has a value very close to $-2.6$, well agrees with the results obtained by \citep{Lipari:2017jou}. On the other hand our result disagrees with previous studies of the CR spectrum below PeV energies that assumed the spectrum to follow an SPL with a soft index between $-2.7$ and $-2.8$ \citep{Abdo_2008}.
While the contribution of unresolved sources to the GDE and its relevant effect on the average spectral index should be taken into account (more discussion on this in the section \ref{sec:conclusions}), yet the spectral index of the GDE measured with HAWC reveals that the CR population producing GDE along the Galactic Plane has {(on average)} spectral index compatible with the spectral index of the locally measured CR population up to tens of TeV {(see section \ref{sec:intro})}.
{The harder spectral indices of the GDE of some sub-regions {(as shown in table \ref{tab:GDE})} are likely due to the contribution of unresolved sources.}

\subsection{Profile Generation} \label{profiles}
{ Galactic longitudinal and latitudinal profiles provide valuable information on the distribution of the GDE along the Galactic Plane.}

{
The longitudinal profile is created by averaging the energy 
flux (see Equation \ref{eq:flux}) within 10 bins, each defined in a 3-degree longitudinal range, over a latitude range of $|b|$  \textless  \xspace  2\degree. The resulting bins collectively span the entire longitudinal range of analysis, covering 30 degrees in total.
Conversely, the latitudinal profile is generated by averaging the energy 
flux over the entire longitude range within 9 bins, each defined in a 1-degree latitudinal range. This approach enabled us to cover the latitude range of $|b|$  \textless  \xspace  4.5\degree.} 

The energy flux (Equation \ref{eq:flux}) is computed over the energy range between 300 GeV {$(E_{min})$} and 100 TeV\footnote{{ In table \ref{tab:GDE}, the energy flux is also computed up to 10 TeV.}} {$(E_{max})$},
\begin{equation}  \label{eq:flux}
\begin{split}
F(E_{min}, E_{max}) & = \int_{E_{min}}^{E_{max}}E \phi(E)dE \\
&\\
& =  \phi_0 \frac{E_0^2}{-\Gamma + 2}   \Bigl( \frac{E}{E_0} \Bigl)^{-\Gamma + 2} \bigg| _{E_{min}}^{E_{max}}
\end{split}
\end{equation}
where $\phi(E) = \phi_0 (E/E_0)^{-\Gamma} $, { as $E_0$ is a reference energy, $\phi_0$ is the differential flux at $E_0$ and $\Gamma$ is the spectral index.}
{
}
Profiles are computed for both the ``original map'' (Figure \ref{sig-org}) and the GDE map (Figure \ref{res_G}); therefore, it is possible to compare the level of the energy flux 

of the total flux measured by HAWC, $F_{tot}$, (red line) with those parameters of the diffuse emission (blue line) along the Galactic Plane.  

In the profiles, vertical error bars represent the statistical error, and horizontal error bars show the width of each bin that GDE parameters are calculated in it.
The brown lines in longitudinal profiles indicate the {sub-}regions as explained in Table \ref{tab:GDE}.

\begin{figure}
  \centering

\includegraphics[width=12cm]{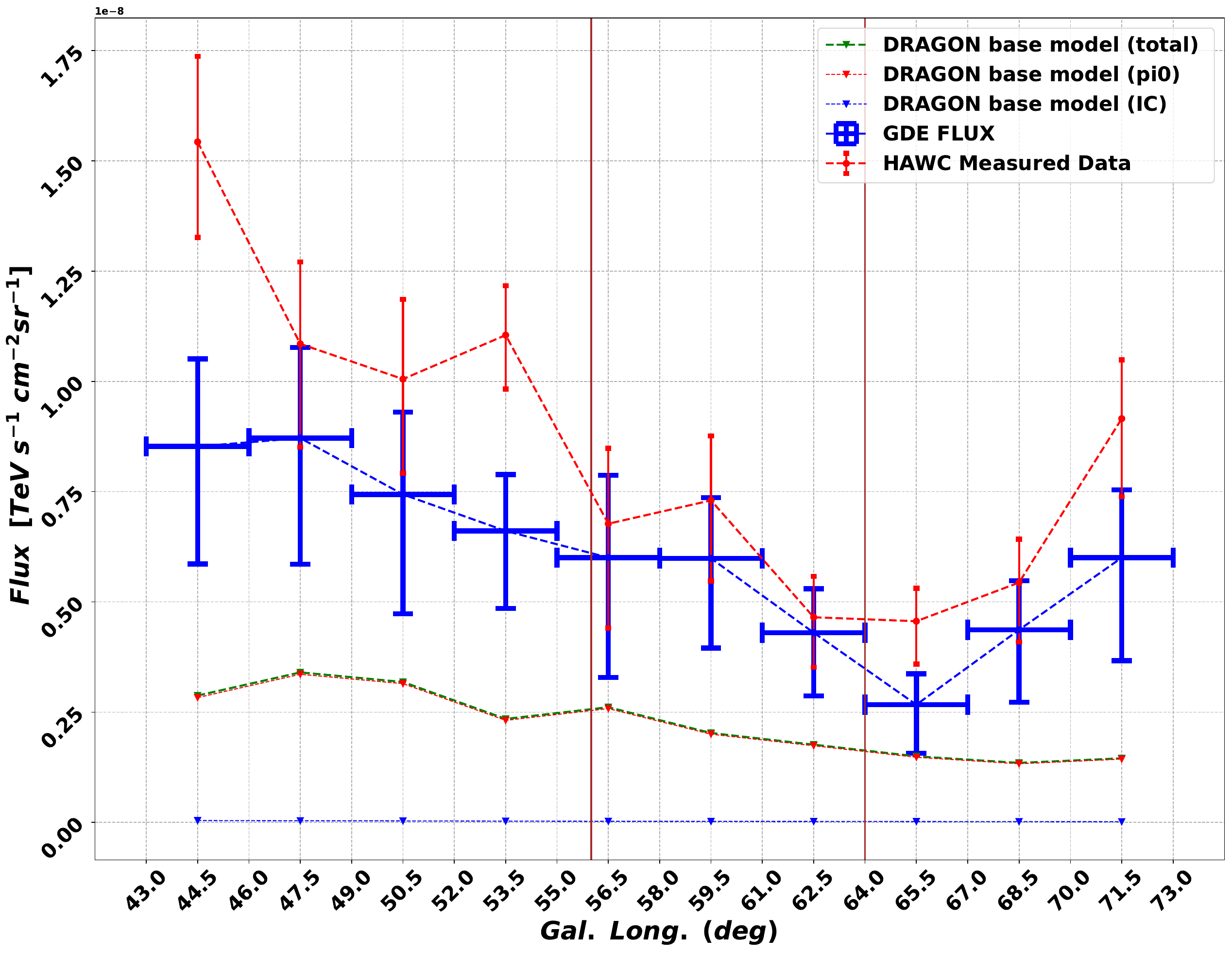}
\caption{Longitude profiles for energy flux between 300 GeV and 100 TeV, the red line refers to the total flux ($F_{tot}$), the blue line represents the GDE, as well as prediction of diffuse emission by \texttt{DRAGON} for $\pi^0$-decay and IC production mechanism.
The brown lines show the border of the regions as explained in Table \ref{tab:GDE}.
}
   \label{lon-profile-flux}
\end{figure}

\begin{figure}
  \centering
\includegraphics[width=12cm]{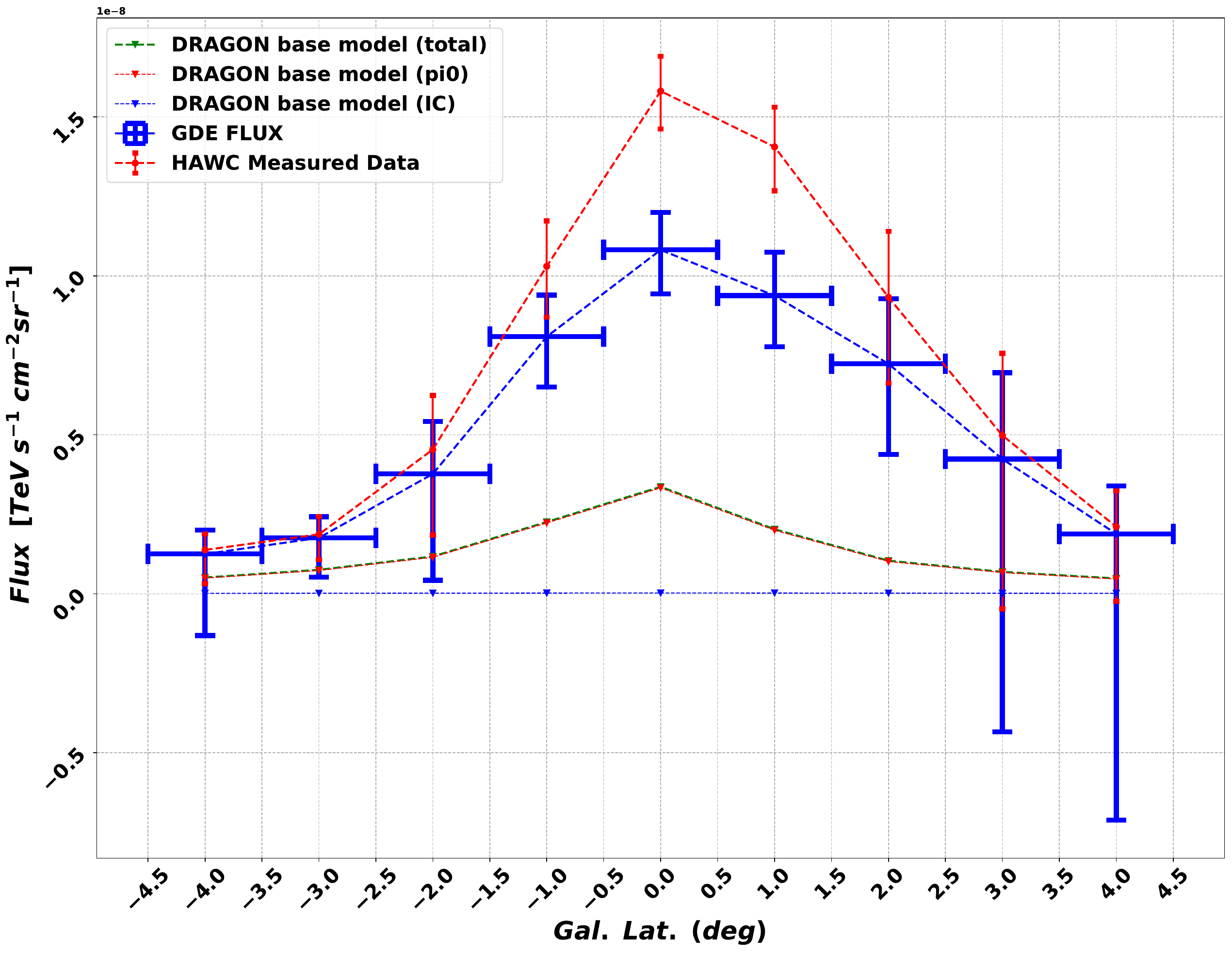}
\caption{Latitudinal profile for the energy flux between 300 GeV and 100 TeV, red line and blue refer to the total flux measured by HAWC, and the \gde respectively. \texttt{DRAGON} estimation for $\pi^0$-decay and IC production mechanism are also presented.}
   \label{lat-profile-flux}
\end{figure}

{In the longitudinal profiles in the range between {56\degree  \textless  \xspace $l$  \textless  \xspace 64\degree} (Figure \ref{lon-profile-flux}), no bright sources were identified in our analysis; the difference between the \gde and the total emission (represented by the blue and red lines, respectively) reaches a minimum value. The f$_{100}$ is 88.1\% (see Table \ref{tab:GDE}) and the spectrum of the total emission is similar to the GDE spectrum.}

Figure \ref{lat-profile-flux} represents the latitude profile of the total energy flux measured by HAWC (red),  \gde flux (blue), and \texttt{DRAGON} estimation for the $\pi^0$-decay, and IC production mechanisms.
The error bars increase on the edge of the latitude profiles due to the lack of statistics in these regions.

\subsection{Comparison with other observations of the GDE}
\begin{figure}[htbp]
\centering
\includegraphics[width=19cm]{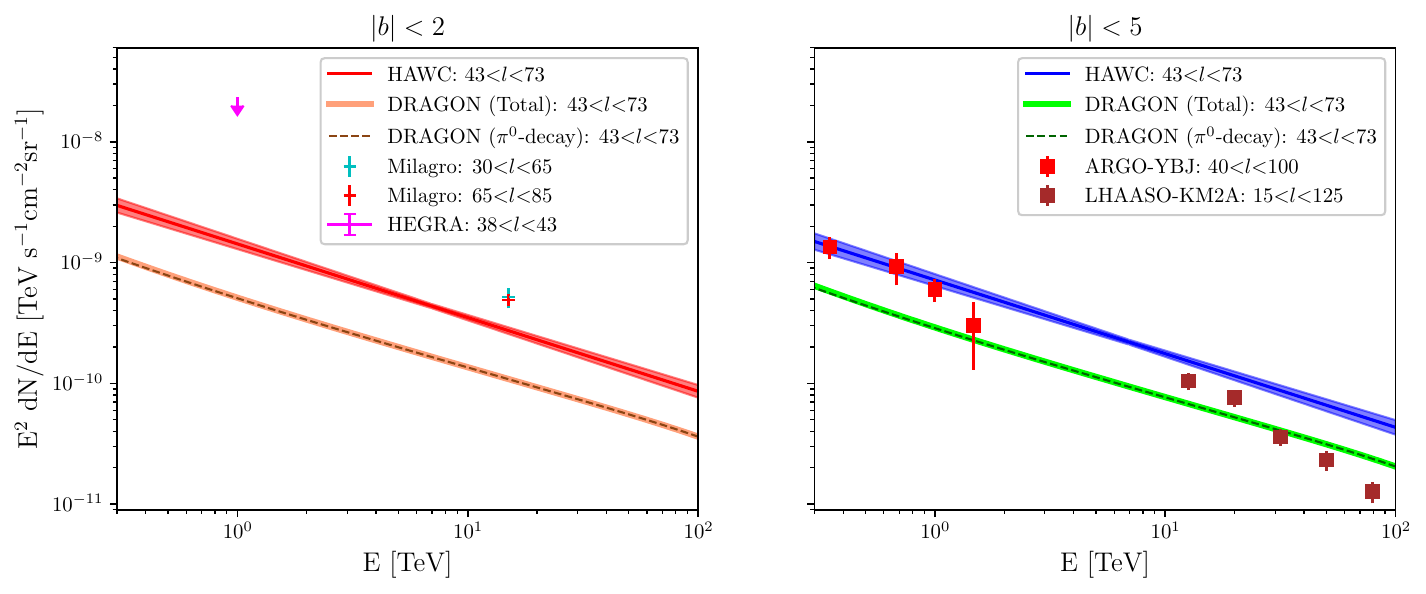}
\caption{
Spectra of the GDE measured by different experiments at different regions, and \texttt{DRAGON} estimations for total and $\pi^0$-decay emission; HAWC and \texttt{DRAGON} within $43^\circ <l<73^\circ$ (left panel: $|b|$ \textless  \xspace  2$^\circ$, and right panel: $|b|$  \textless  \xspace  5$^\circ$),  { statistical errors and the systematic errors are listed in Table \ref{tab:GDE}}.
Milagro at 15 TeV for 2 regions within $|b|$  \textless  \xspace  2\degree  \citep{Abdo_2008}. An upper limit quoted by HEGRA-IACT (99\%  confidence level) in $|b|$  \textless  \xspace  2$^\circ$ \citep{2001A&A...375.1008A}.
ARGO-YBJ in $40^\circ <l<100^\circ$ \citep{Bartoli_2015}, and LHAASO-KM2A in $15^\circ <l<125^\circ$ \citep{cao2023measurement}, both within $|b|$  \textless  \xspace  5$^\circ$.} 
\label{fig:spect}
\end{figure}

The observations of TeV Galactic diffuse gamma-ray emission, performed by different experiments such as Milagro \citep{Abdo_2008}, LHAASO-KM2A \citep{cao2023measurement},
HEGRA-IACT \citep{2001A&A...375.1008A} and ARGO-YBJ \citep{Bartoli_2015} are shown in Figure \ref{fig:spect}.
The Milagro points for longitudes $30^\circ <l<65^\circ$ and $65^\circ <l<85^\circ$ are obtained at 15 TeV \citep{Abdo_2008} energies. An upper limit was set by HEGRA-IACT at 1 TeV with 99\% confidence level (C.L.) \citep{2001A&A...375.1008A}.\\
 
{ When compared with the results obtained with Milagro, the findings of this study reveal a smaller level of the \gde as measured by HAWC data.
This disparity can be attributed to the utilization of a model (see section \ref{process}) that more effectively accounts for source emissions, leading to a better fit of the data and improved estimation of the true level of \gde.
As a consequence, the HAWC study reaches a lower threshold for the \textit{true} \gde compared to the previous work by Milagro.
LHAASO reported the energy spectrum of the \gde within the energy range of 10 TeV to 1 PeV. The observed spectrum, found in the inner region ($15^\circ <l<125^\circ$, $|b|$  \textless  \xspace  5$^\circ$), follows a power-law function with an index of -2.99 $\pm$ 0.04.

}

HEGRA-IACT (Figure \ref{fig:spect}) reported an upper limit for GDE above 1 TeV   (99\% C.L.)  with a presumed spectral index of -2.6 \citep{2001A&A...375.1008A}.
As shown in Figure \ref{fig:spect}, the HAWC measurement is below HEGRA's upper limit; this excess suggests a significant contribution of unresolved sources in GDE measured by HEGRA-IACT, likely due to the non-ideal modeling for the sources in the HEGRA-IACT analysis. 
Moreover, HEGRA-IACT has measured the \gde in a  narrow region closer to the Galactic center (lower Galactic longitude), which is expected to have a more significant diffuse emission (as it can be seen in Figure \ref{res_G}).
In Figure \ref{fig:spect}, the estimated spectra of \texttt{DRAGON} for the $\pi^0$-decay mechanism, and total diffuse emission (which is a sum of  $\pi^0$-decay and IC) are shown. 
The IC contribution modelled with \texttt{DRAGON} is negligible. In Figure \ref{fig:spect} we also include a comparison with the ARGO-YBJ measurements. However, we remark that a comparison of HAWC findings with the ARGO-YBJ results is difficult. ARGO-YBJ reported the GDE emission from  higher longitudes (further away from the Galactic center), with energies ranging from $\sim$ 350 GeV to $\sim$ 2 TeV.

\section{Discussion and Conclusion}
\label{sec:conclusions}

We have presented the first analysis of the spectral and angular distribution of the diffuse gamma-ray emission measured by HAWC above 1 TeV over a portion of the Galactic Plane between longitude and latitude of $l\in[43^\circ,73^\circ]$ and  $b\in[-5^\circ,5^\circ]$, respectively. We have determined both the longitudinal and latitudinal profiles of this emission and its spectrum in several sub-regions. 

We have found that the spectrum of the emission is well fitted by an SPL model with index $-2.61 \pm 0.03$ (see table \ref{tab:GDE}). Such a spectral index well agrees with the emission being generated by population of background CR protons and heavier nuclei, whose spectral shape very closely mimics the spectral shape of the CR spectrum probed by experiments near the Earth (see section \ref{sec:intro}). Our results support the picture in which the CR spectral hardening found by PAMELA and AMS at the rigidity of about 300 GV is a large-scale feature as also suggested by the AMS measurements of secondary CR nuclei spectra \citep{Aguilar:2018njt}. 

{ In this study, we compared the predictions of a reference model, known as the \textit{base model}, implemented using the \texttt{DRAGON} code, with the measured spectral shape of the \gde in the sky-window observed by the HAWC. The base DRAGON model solves CR propagation in the Galaxy assuming standard CR transport properties, namely a CR diffusion coefficient as obtained from measurements of secondary CR ratios. The CR spectra over the whole Galaxy are then convolved with the gas distribution. The spectral properties of the \texttt{DRAGON} \textit{base model} agreed well with the \gde measurements.  The model under-predicted the \gde measured by HAWC by a factor of $\sim$2, averaged over the entire region (see Figure \ref{fig:spect}).}

{This discrepancy may be attributed to a variety of factors, including potential underestimation of the density of CRs or gas, as well as uncertainties related to their respective distributions, particularly in the case of CRs which are still largely unknown. 
In addition, the contribution from unresolved sources cannot be overlooked, as it has been shown that up to 90$\%$ of the diffuse radiation at TeV energies can originate from such sources \citep{2020ApJ...904...85C,2020A&A...643A.137S}. 
Another limitation that cannot be ignored is the non-ideal model for sources used in this study, which causes the GDE to include emission from sources left in the source-subtracted map.
Furthermore, the TeV halos of pulsar wind PWNe are also expected to make a significant contribution to the emission at both GeV and TeV energies \citep{2022CmPhy...5..161V}.
In this sense, our measurement of the GDE can be seen as an upper limit to the \textit{truly} diffuse emission. 
Also, it should be kept in mind that systematic errors (angular and energy resolution) restrict the analysis precision  \citep{Abeysekara_2017}.

Future HAWC analyses using higher-quality data, such as Pass 5 data, will be necessary to further elucidate the difference between the measured and predicted diffuse fluxes. Very recently LHAASO published a measurement of the diffuse emission from a large region of the Galactic Plane, 15\degree  \textless  \xspace $l$  \textless  \xspace 125\degree and -5\degree  \textless  \xspace $b$  \textless  \xspace 5\degree, by excluding large portions of the Galactic Plane, and reported a significantly lower level for the diffuse emission with respect to HAWC \citep{2023arXiv230505372C}. 
The analysis of TeV diffuse emission will be a major scientific objective for future gamma-ray observatories, such as the SWGO observatory \citep{abreu2019southern} and the Cherenkov Telescope Array \citep{consortium2017science}, which will benefit from improved sensitivity and angular resolution.}

\appendix
\label{sec:appendix}

\section{Differential Flux Profiles} \label{sec:DFP}
{ As formerly discussed in section \ref{profiles}, longitudinal and latitudinal profiles are used to study the distribution of \gde along the Galactic Plane. 
This appendix presents the longitudinal and latitudinal profiles for the differential flux of the GDE emission at E$_0$ = 7 TeV. The longitudinal profile is obtained by averaging the differential flux over a range in latitude $|b|$ \textless \xspace 2\degree. 
The latitudinal profile is obtained by averaging the differential flux over a range in longitude $l \in [43^\circ,73^\circ]$, which covers the latitude range of $|b|$ \textless \xspace 4.5\degree.}

\begin{figure}
  \centering
  \includegraphics[width=12cm]{ 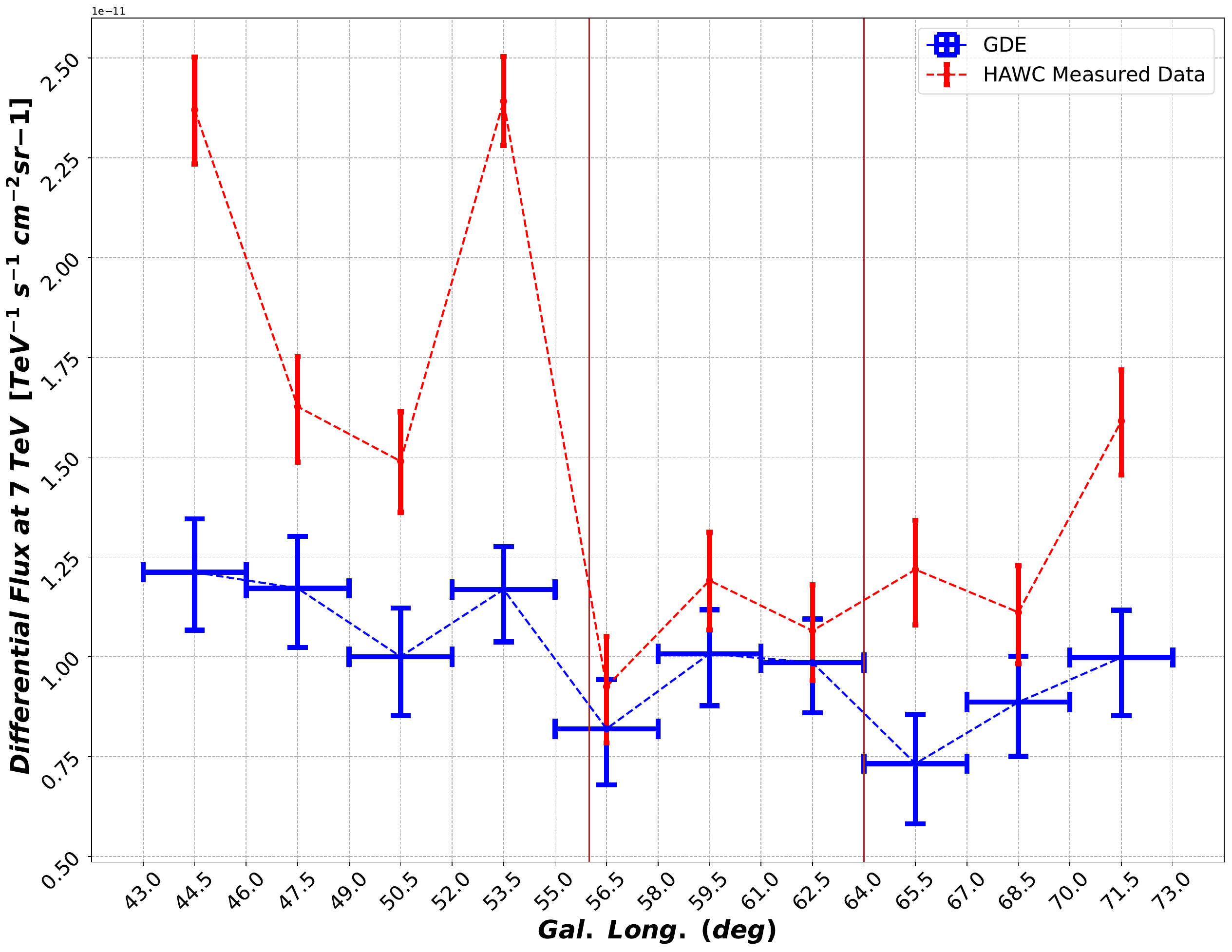}
\caption{Longitudinal profile of the  {differential flux} (at 7 TeV); the blue line represents the  { \gde differential flux} with error bar, while the red line shows the longitudinal profile of the total flux ($F_{tot}$) measured by HAWC. The vertical brown lines are borders of sub-regions.}

\label{lon-profile-norm}
\end{figure}

\begin{figure}
\centering 
 \includegraphics[width=12cm]{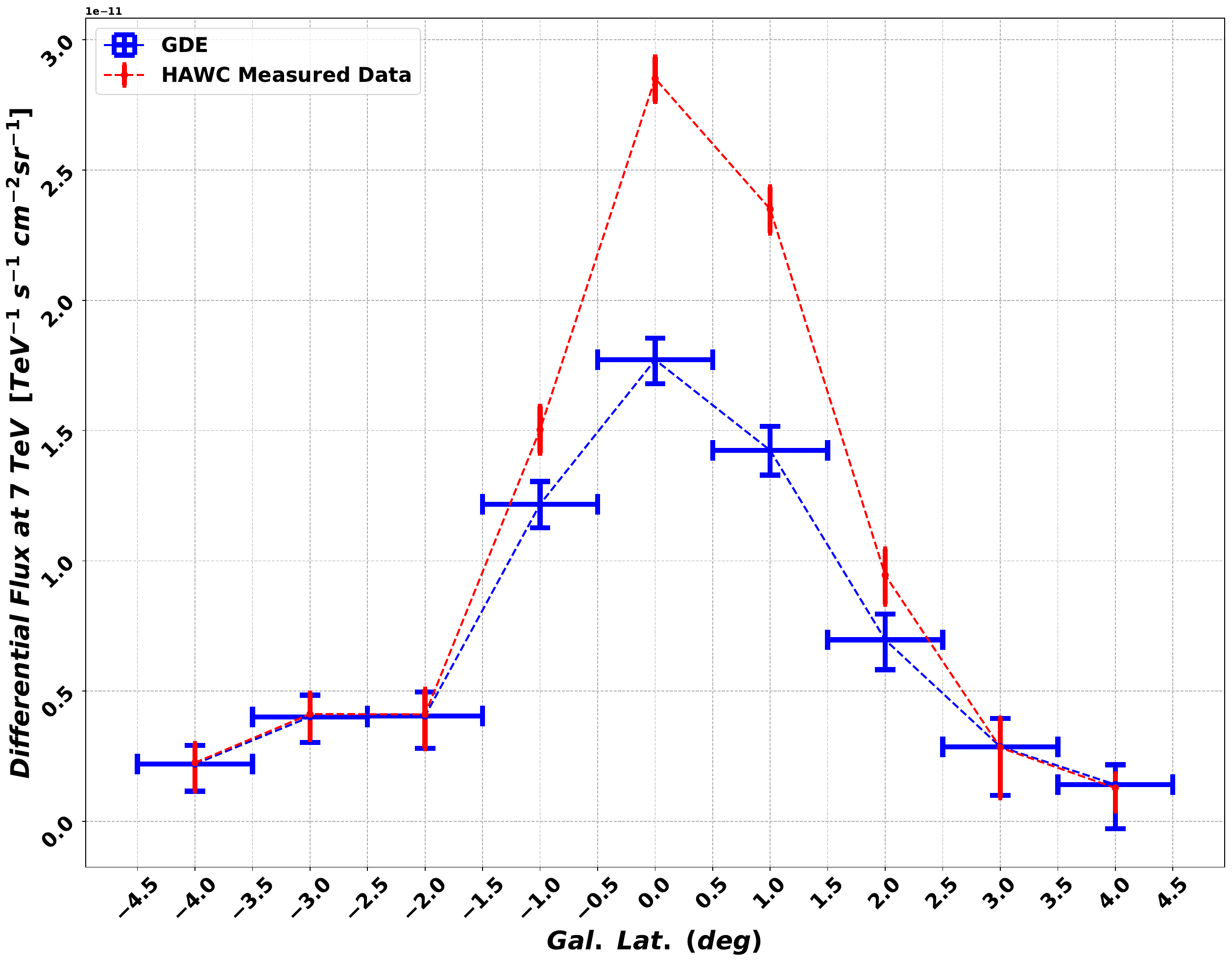}
\caption{Latitudinal profile of the {differential flux} (at 7 TeV). The {differential flux} of the diffuse emission is represented by blue lines with error bars while the red line refers to the {total differential flux} measured by HAWC.} 
   \label{lat-profile-norm}
\end{figure}

{ 
The longitudinal profile of the differential flux (Figure \ref{lon-profile-norm}), as discussed in section \ref{profiles}, exhibits a minimum difference between the \gde and total emission, represented by the blue and red lines, respectively, in the range of {56\degree  \textless  \xspace $l$  \textless  \xspace 64\degree}. 
In addition, Figure \ref{lat-profile-norm} shows the latitudinal profile of the differential flux of the GDE and the total emission measured by HAWC (within the same bins).
}

\section{Comparison with ARGO-YBJ measurements} \label{ARGO-DRAGON}

\begin{figure}
\centering
\includegraphics[width=12cm]{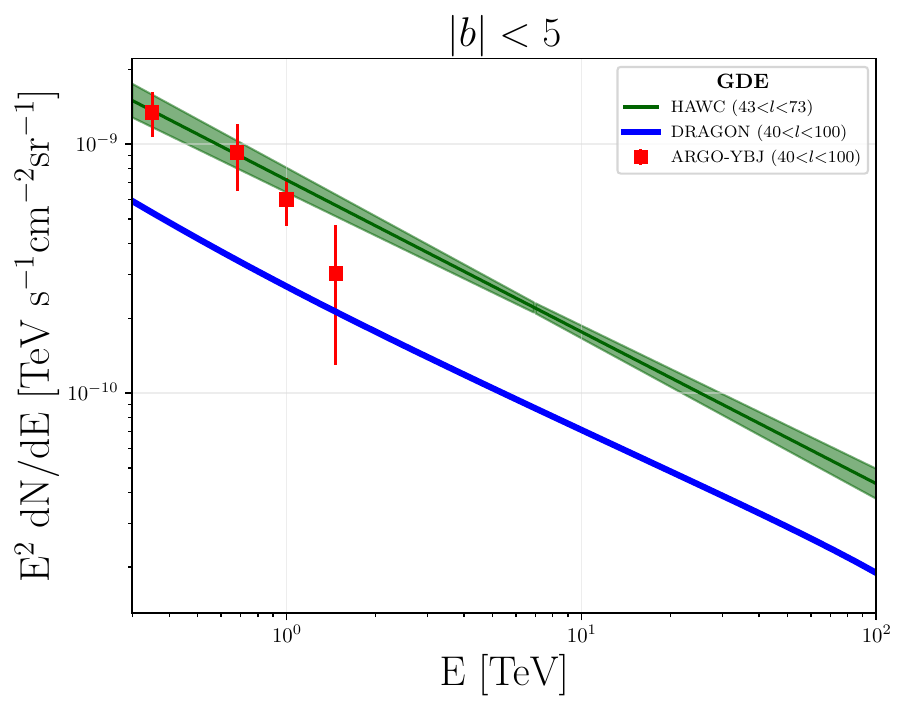}
\caption{Spectra of the GDE measured by HAWC and ARGO-YBJ in $|b|$ \textless \xspace 5\degree and in $43^\circ <l<73^\circ$ and $40^\circ <l<100^\circ$, respectively \citep{Bartoli_2015} and estimated with \texttt{DRAGON} within $|b|$  \textless \xspace 5$^\circ$ and $40^\circ <l<100^\circ$.}
\label{ARGO:points}
\end{figure}

{The measurement of the diffuse emission by ARGO-YBJ \citep{Bartoli_2015} at Galactic longitudes $40^\circ <l<100^\circ$ and Galactic latitudes $|b|$ \textless  \xspace 5\degree and for an energy range from $\sim$350 GeV to $\sim$2 TeV is shown in Figure \ref{ARGO:points}. 
The measured flux at three median energies, 350 GeV, 680 GeV, and 1.47 TeV (with uncertainties of about 30\%) is above the \textit{DRAGON} estimations of GDE.  Additional flux from unresolved sources and from PWNe halos} might explain the discrepancy
between the ARGO-YBJ results and the \textit{DRAGON} predictions. The spectral index of the GDE as measured by ARGO-YBJ is -2.9$\pm$0.31, which is softer than HAWC spectral index. For comparison, we show also the HAWC flux measured in $43^\circ <l<73^\circ$ and Galactic latitudes $|b|$ \textless  \xspace 5\degree, which is however measured from an inner region of the Galactic Plane. 

\section{ \texttt{DRAGON}} \label{sec:dragon}
The \texttt{DRAGON} code \citep{Evoli_2008,Evoli_2017} solves the CR transport equation accounting for diffusion, re-acceleration, advection, energy losses and spallation onto the interstellar gas. 
A conventional approach in which the diffusion coefficient is assumed to be isotropic in the whole Galaxy and homogeneous in the Galactic Plane is chosen here. The reference set-up (\textit{base model}) is presented in \citep{Fornieri:2019ddi} where it was shown to reproduce the CR proton, helium and electron spectra measured by AMS and, at larger energies relevant here, CREAM \citep{Yoon_2017} as well as CALET \citep{Adriani:2019aft} results. 

As mentioned in the introduction, this requires a spectral hardening in the primary nuclei spectra at an energy $\sim 300\ {\rm GeV}/{\rm n}$. 
For the same model, \texttt{DRAGON} predicts both the spatial and energy distribution in the whole Galaxy for all relevant CR species which are, then, convolved with the proper cross sections and target gas/radiation distributions to determine the line-of-sight integrated gamma-ray fluxes due to $\pi^0$ decay, IC and bremsstrahlung at each point of the sky.  \par
For the H$_2$ and H$_{\rm I}$, H$_{\rm II}$ and helium gas components the same distributions adopted in the \textit{GALPROP} code \citep{Vladimirov_2011} are assumed. The H$_2$ is based on the observed CO emission maps in several Galactocentric rings \citep{1988ApJ...324..248B} after multiplication by a radial dependent CO-to-H$_2$ conversion factor which, following \citep{Gaggero_2015xyz}, is assumed to be $1.9(5) \times 10^{20}~({\rm cm}^{2}{\rm K}~{\rm km}~{\rm s}^{-1})^{-1}$ for Galactocentric radii smaller (larger) than 7.5 kpc. 
The lower value is in agreement with several astrophysical measurements \citep{Bolatto:2013ks} while the larger value at large radii provides an effective compensation of the otherwise too steep longitude profile of the gamma-ray diffuse emission (\textit{gradient problem}) as assumed in other related works (see {\em e.g.} \citep{1996A&A...308L..21S}).  \\
The \textit{base model} is tuned to reproduce the morphology and 
the spectrum of the diffuse emission 
originated by CRs as measured by AMS in the sky-window under consideration.

\bibliography{sample63}{}
\bibliographystyle{aasjournal}

\end{document}